\documentclass[pra,aps,nopacs,superscriptaddress,nofootinbib,longbibliography,notitlepage,twocolumn]{revtex4-2}

\usepackage{amsmath}  \usepackage{amssymb}  \usepackage{amsfonts}  \usepackage{bm}  \usepackage{bbm}   \usepackage{braket}  \usepackage{color}  \usepackage{comment}  \usepackage{dcolumn}  \usepackage{enumerate}  \usepackage{epsfig}  \usepackage{gensymb}  \usepackage{graphicx}  \usepackage{indentfirst}  \usepackage{lmodern}  \usepackage{mathrsfs}  \usepackage{mathtools}  \usepackage{psfrag}  \usepackage{pst-all}  \usepackage{soul}   \usepackage{xcolor}
\usepackage{float}
\usepackage[colorlinks,linkcolor=blue,citecolor=blue,urlcolor=blue,hyperindex,driverfallback=dvipdfm]{hyperref}  \usepackage[T1]{fontenc}
\usepackage{makecell}
\usepackage{orcidlink}

\def\ii{{\rm i}}  \def\ee{{\rm e}}
  \def\kB{{k_{\rm B}}}
  
        \def\Eb{{\bf E}}      \def\fb{{\bf f}}          \def\Jb{{\bf J}}            \def\Pb{{\bf P}}  \def\pb{{\bf p}}      \def\Rb{{\bf R}}  \def\rb{{\bf r}}      \def\vb{{\bf v}} 
\def\xx{\hat{\bf x}}  \def\yy{\hat{\bf y}}  \def\zz{\hat{\bf z}}    \def\rr{\hat{\bf r}}      \def\RR{\hat{\bf R}}  
\def\EF{{E_{\rm F}}}  \def\vF{v_{\rm F}}   
      
\def\mydeg{^{\circ}} 
\def\vtheta{\vec{\theta}}
\def\calE{\mathcal{E}} \def\vcalE{\vec{\mathcal{E}}}
\def\q{Q} \def\j{m}

\usepackage{scalerel}
\def\shrinkage{-2.4mu}
\def\vecsign#1{\rule[1.388\LMex]{\dimexpr#1-2.5pt}{.36\LMpt}%
  \kern-6.0\LMpt\mathchar"017E}
\def\dvecsign#1{\rule{0pt}{7\LMpt}\smash{\stackon[-1.989\LMpt]{%
  \SavedStyle\mkern-\shrinkage\vecsign{#1}}%
  {\rotatebox{180}{$\SavedStyle\mkern-\shrinkage\vecsign{#1}$}}}}
\def\dvec#1{\ThisStyle{\setbox0=\hbox{$\SavedStyle#1$}%
  \def\useanchorwidth{T}\stackon[-4.2\LMpt]{\SavedStyle#1}{\,\dvecsign{\wd0}}}}
\usepackage{stackengine,amsmath}
\stackMath

\usepackage{array,multirow}
\newcolumntype{P}[1]{>{\centering\arraybackslash}m{#1}}

\begin{document}
\def\bibsection{\section*{\refname}}

\title{Active steering of cathodoluminescence through a generalized Smith-Purcell effect
}

\author{Eduardo~J.~C.~Dias\,\orcidlink{0000-0002-6347-5631}}
\email[Eduardo~J.~C.~Dias: ]{dias@mci.sdu.dk}
\affiliation{POLIMA---Center for Polariton-driven Light--Matter Interactions, University of Southern Denmark, Campusvej 55, DK-5230 Odense M, Denmark}

\author{A.~Rodr\'{\i}guez~Echarri\,\orcidlink{0000-0003-4634-985X}}
\affiliation{Center for Nanophotonics, NWO Institute AMOLF, 1098 XG Amsterdam, The Netherlands}
\affiliation{Max-Born-Institut, 12489 Berlin, Germany}

\author{Theis~P.~Rasmussen\,\orcidlink{0000-0001-6347-0788}}
\affiliation{POLIMA---Center for Polariton-driven Light--Matter Interactions, University of Southern Denmark, Campusvej 55, DK-5230 Odense M, Denmark}

\author{F.~Javier~Garc\'{\i}a~de~Abajo\,\orcidlink{0000-0002-4970-4565}}
\affiliation{ICFO-Institut de Ciencies Fotoniques, The Barcelona Institute of Science and Technology, 08860 Castelldefels (Barcelona), Spain}
\affiliation{ICREA-Instituci\'o Catalana de Recerca i Estudis Avan\c{c}ats, Passeig Llu\'{\i}s Companys 23, 08010 Barcelona, Spain}

\author{Joel~D.~Cox\,\orcidlink{0000-0002-5954-6038}}
\email[Joel~D.~Cox: ]{cox@mci.sdu.dk}
\affiliation{POLIMA---Center for Polariton-driven Light--Matter Interactions, University of Southern Denmark, Campusvej 55, DK-5230 Odense M, Denmark}
\affiliation{Danish Institute for Advanced Study, University of Southern Denmark, Campusvej 55, DK-5230 Odense M, Denmark}

\begin{abstract}
    Optical metasurfaces can shape the near fields of energetic electrons, enabling Smith-Purcell (SP) emission. We introduce a generalized SP effect relying on finite periodic arrays whose elements possess individually tunable polarizabilities, allowing us to explore higher-order SP radiation. By controlling the amplitude and phase of each of the elements, we show through rigorous theory the ability to create an SP steering device. In particular, we explore the active tuning capabilities of doped graphene, and thermally driven phase-change materials, which we compare with standard passive plasmonic structures made of gold and silver. Our results establish programmable electron-driven light sources and spectroscopic probes spanning the terahertz-to-visible range, advancing tunable metasurfaces for next-generation electron-photon technologies.
\end{abstract}

\maketitle
\date{\today}

\section{Introduction}

Optical metasurfaces have recently emerged as two-dimensional (2D) platforms for developing compact devices capable of manipulating light at the nanoscale~\cite{yu2014flat, hu2021review}. The functionalities of metasurfaces are rich and diverse, including nonlinear frequency conversion~\cite{li2017nonlinear, keren2018shaping, semmlinger2018vacuum}, optical holography~\cite{wan2017metasurface, deng2017metasurface}, and wavefront shaping~\cite{karimi2014generating}, just to mention a few. The foundation of these various applications builds upon periodic arrangements of subwavelength nanostructures, engineered such that the localized excitation of all the individual constituents converges to shape an overall targeted scattering response. Furthermore, the combination of the 2D metasurfaces---being easier to fabricate than their 3D metamaterial counterparts---along with a typically lower degree of losses, endows these planar architectures with unique advantages for many applications in nanophotonics.

While metasurfaces are conventionally used to manipulate light impinging from the far-field~\cite{wang2018broadband}, the control of optical near-fields by metasurfaces has been demonstrated both theoretically and experimentally~\cite{deshpande2018direct, li2020collective}. In this context, energetic free electrons serve as an excellent source of broadband evanescent electromagnetic fields, which can be directed over a metasurface with exceptional spatial precision~\cite{garciadeabajo2010optical}. Specifically, the optical excitation of elements in a structured surface by swift electrons can generate light emission covering a vast range of frequencies, thus holding significant promise for applications using light sources at the nanoscale~\cite{adamo2009light, rosolen2018metasurface}. In particular, free-electron induced Smith-Purcell (SP) radiation~\cite{smith1953visible}, in which the time-varying Coulomb field of a charged particle moving parallel to a periodically structured surface interacts with optical modes of the surrounding material~\cite{su2019manipulating,paper014}, has been intensely explored in many systems, including Babinet surfaces~\cite{wang2016manipulating}, arrays of noninteracting particles~\cite{garaev2021theory}, and chirped metagratings~\cite{karnieli2022cylindrical}.

Smith-Purcell (SP) radiation depends crucially on both the geometrical and intrinsic material properties of a metasurface. While the geometry of metal gratings supporting collective free electron excitations is typically controlled passively (e.g., by changing the size of individual elements and their periodicity), intrinsic material parameters, such as dielectric permittivity, are rather challenging to control. Further customization of SP emission can be obtained by relaxing stringent conditions on the periodicity of an array, with aperiodic structures providing more complex far-field emission patterns~\cite{saavedra2016smith} and near-field focusing~\cite{karnieli2022cylindrical}. However, most systems explored in this context lack the ability to actively control the emission properties of SP radiation, such as directionality, far-field amplitude, and polarization.

In this work, we present a generalized study of SP radiation in finite-size periodic arrays with individually tunable polarizable elements. Opening with a summary of conventional SP radiation in periodic arrays of identical scatterers, we generalize the formalism to investigate the far-field emission characteristics of finite arrays of arbitrary dipoles, emphasizing the coupling to discrete emission channels with specific directionality. The generalized SP formalism is then applied to steer the cathodoluminescence (CL) produced by an electron passing over a periodic array of non-uniform elements by customizing the polarizability of each scatterer. We explore this concept by simulating generalized SP emission in paradigmatic actively tunable nanophotonic architectures: a periodic graphene nanoribbon array, supporting electrically tunable plasmon resonances, and particles comprised of phase-change materials that can be optically activated through variations in temperature. Our results open new avenues for electron spectroscopy, light sources, photon-electron interactions, and optimizing the inverse Smith-Purcell effect used in laser-driven linear accelerators.

\section{Results}

\subsection{Generalized Smith-Purcell condition}

\begin{figure*}[htbp]
    \centering
    \includegraphics[width=0.9\linewidth]{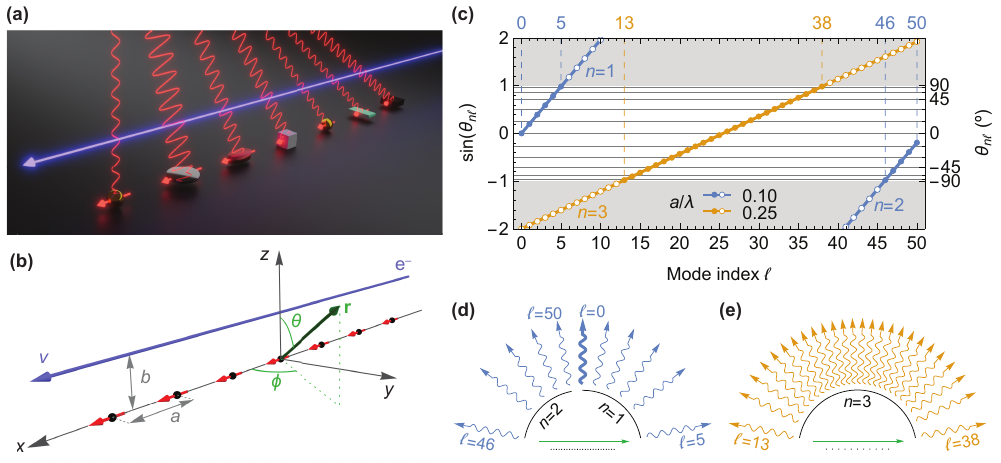}
    \caption{\textbf{Generalized Smith-Purcell emission.} \textbf{(a)} Artistic sketch of the considered system, composed of a periodic array of scatterers interacting with an electron beam moving parallel to the array. \textbf{(b)} Simplified scheme of the sketch in (a), where we indicate the period $a$ of the array, the electron (e$^-$) velocity $v$ and impact parameter $b$. We further indicate the adopted coordinate system and the polar and azimuthal angles $\theta$ and $\phi$ for a general position vector $\rb$. All scatterers are taken as point dipoles with dipole moments indicated by red arrows. \textbf{(c)} Generalized Smith-Purcell (GSP) emission condition in the $\phi=0$ plane, given by Eq.~\eqref{eq:SPcondition}, for an electron with velocity $v=0.1\,c$ passing near an array with $N=51$ elements and period indicated by the legend. The SP order $n$ for each curve is marked next to it, while all remaining values of $n$ lie outside the plot region for the considered values of $a/\lambda$. The gray areas indicate the regions where $|\sin \theta_{n\ell}|> 1$ and therefore constructive interference is kinematically forbidden. Dots inside (outside) this region are marked as unfilled (filled), and the corresponding emission angle $\theta_{n\ell}$, when allowed, is denoted along the right axis. \textbf{(d,e)} Schematic illustrations of the available GSP emission channels for each of the color-coordinated arrays in (c). Each arrow corresponds to a specific $n,\ell$ pair (see labels) and is represented outgoing at the $\theta_{n\ell}$ angle. In (d), the $\ell=0$ case is highlighted by a thicker arrow, denoting the conventional SP emission of a uniform array.}
    \label{fig1}
\end{figure*}

We start by considering a general one-dimensional array composed of $N$ elements periodically placed at coordinates $\rb_j=ja\xx$ ($j=0,\cdots,N-1$) along the $x$ axis, where $a$ is the lattice period, as depicted in Figs.~\ref{fig1}(a,b). The array is surrounded by vacuum and is excited by a swift electron passing at a distance $b>0$ and moving with velocity $\vb=v\xx$ that generates at frequency $\omega$ (i.e., wavelength $\lambda=2\pi c/\omega$) an external field $\Eb^{\rm ext}(\rb_j)$ given by Eq.~\eqref{eq:Eext} in Methods. In general, we assume that each element $j$ in the array can exhibit a distinct polarizability response $\alpha_j$, and thus will generate a distinct dipole moment
\begin{align}
\pb_j=\alpha_j\cdot\Big[\Eb^{\rm ext}(\rb_j)+\sum_{i\neq j} \mathcal{G}_{ji}\cdot \pb_i\Big],
\label{eq:pj}
\end{align}
where $k=\omega/c=2\pi/\lambda$ is the free space optical wave vector and the term $\mathcal{G}_{ji}=(k^2+\nabla \otimes \nabla)\ee^{\ii k r_{ji}}/r_{ji}$ represents the dipole-dipole interaction between array elements $i$ and $j$ in terms of their distance $r_{ji}=|\rb_j-\rb_i|$. The self-consistent equation above can be inverted to find the induced dipole moment in each array element, which we can write in the form $\pb_j=\pb^0_j \ee^{\ii \omega j a/v}$, anticipating from Eq.~\eqref{eq:Eext} that the time delay associated with the finite electron velocity introduces a phase difference of $\omega a/v$ between two consecutive dipole elements. In general, each dipole $\pb_j^0$ may contain components along $x$ and $z$, but components along $y$ are forbidden by symmetry.

The far-field cathodoluminescence (CL) emission by such an array, as a function of emission direction $\rr=(\sin\theta\cos\phi,\sin\theta\sin\phi,\cos\theta)$, i.e., along polar and azimuthal angles $\theta$ and $\phi$, respectively, represented in Fig.~\ref{fig1}(b), can be constructed by the superposition of the far-field emitted by each array element,
\begin{align}
  \fb(\rr) = k^2 (1-\rr\otimes\rr) \cdot \sum_j \pb_j^0 \ee^{\ii k j a/\beta} \ee^{-\ii k (\rb_j \cdot \rr)},
  \label{eq:SPf}
\end{align}
where $\beta=v/c$ and we introduce a phase $k (\rb_j \cdot \rr)=k a j \sin\theta\cos\phi$ to account for the optical path difference between the field emitted by two consecutive array elements.

In traditional SP emission, an infinite array of identical elements is considered, for which we find $\pb_j^0=\pb^0$ for all values of $j$. Under such conditions, the far-field intensity is nonvanishing only if the combined phases in Eq.~\eqref{eq:SPf} sum to an integer multiple of $2\pi$, which, at the $xz$ plane defined by $\phi=0$ (or $2\pi$), leads to the usual SP condition,
\begin{align}
  \sin \theta_n = \frac{1}{\beta}-\frac{n\lambda}{a},
  \label{eq:SP0condition}
\end{align}
where $n\in \mathbb{Z}$ denotes the SP order. At planes where $\phi\neq 0$, the right-hand side of Eq.~\eqref{eq:SP0condition} needs to be modified by a global $1/\cos\phi$ factor that shifts the SP emission angle $\theta_n$ provided that $|\sin\theta_n|$ remains smaller than 1.

Equation~\eqref{eq:SP0condition} holds only when the induced dipole is uniform across all array elements, which is satisfied if the array is infinite and all its elements are equivalent, but breaks down if one of those conditions is not met. Instead, in the general case where the dipoles respond non-uniformly to the external field (and we are no longer able to remove the terms $\pb_j^0$ from the $j$-sum), it is convenient to introduce the Fourier decomposition of the induced dipole array, $\pb_{j}^0 = \sum_{\ell} \tilde{\pb}_{\ell} \ee^{2\pi \ii j \ell/N}$, which allows us to rewrite the sum in Eq.~\eqref{eq:SPf} as 
\begin{align}
  \sum_j \pb_j^0 \ee^{\ii k j a/\beta} \ee^{-\ii k j a (\rr \cdot \xx)} = \sum_{\ell} \tilde{\pb}_{\ell} \sum_j \ee^{\ii j \chi_{\ell}},
  \label{eq:jsump}
\end{align}
with $\chi_{\ell}=2\pi \ell/N + k a(1/\beta- \rr \cdot \xx)$. By doing so, we are able to decouple the dipole moment (now written in terms of the harmonic array modes denoted by $\ell$) from the array positions (denoted by $j$), which, crucially, allows us to analytically evaluate the $j$-sum in Eq.~\eqref{eq:jsump} as 
\begin{align}
 \sum_j \ee^{\ii j \chi_{\ell}}=\begin{cases}
  \ee^{\ii (N-1) \chi_{\ell}/2}\dfrac{\sin(N \chi_{\ell}/2)}{\sin(\chi_{\ell}/2)}, & \chi_{\ell} \neq 2\pi n, \\
  N, & \chi_{\ell}=2\pi n,
  \end{cases}
  \label{eq:chiell}
\end{align}
where, as above, we take $n$ to be an integer number. In this case, the far-field angular distribution can be readily found by combining Eqs.~\eqref{eq:SPf}, \eqref{eq:jsump}, and \eqref{eq:chiell}, and exhibits a resonance signaled by the condition $\chi_{\ell}=2\pi n$, which, in the plane defined by $\phi=0$, leads to the generalized Smith-Purcell (GSP) condition
\begin{align}
    \sin \theta_{n\ell} = \frac{1}{\beta}-\left(n-\frac{\ell}{N}\right)\frac{\lambda}{a}.
\label{eq:SPcondition}
\end{align}
This condition is modified with respect to the standard SP term by the emergence of a new parameter $\ell=0,\cdots,N-1$. When all dipole elements in the array are equally polarized, only the term $\ell=0$ survives upon evaluating Eq.~\eqref{eq:pell}, which reconciles Eqs.~\eqref{eq:SP0condition} and \eqref{eq:SPcondition} and reveals that traditional SP is the particular case of GSP arising under such conditions (with all other channels unavailable). However, when the response of the array is non uniform, Eq.~\eqref{eq:SPcondition} reveals that new channels emerge as possible directions of constructive interference CL emission for each combination of $n$ and $\ell$ that fulfills $|\sin \theta_{n\ell}|\leq 1$, as controlled by the system parameters $\beta = v/c$, $a/\lambda$, and $N$.

In Fig.~\ref{fig1}(c), we graphically represent Equation~\eqref{eq:SPcondition} for the case of an array with $N=51$ elements and an electron with velocity $v=0.1\,c$ ($\approx 2.6$~keV). Two different values of the array $a/\lambda$ are considered, as marked in the respective legend, with respect to some arbitrary design wavelength $\lambda$. In blue, we see an array with $a/\lambda=0.1$, which yields a standard SP emission angle ($\ell=0$) of $\theta_{1,0}=0$\degree, which is represented by the thicker arrow in Fig.~\ref{fig1}(d). However, for a non-uniform array, the GSP emission analysis of this system reveals that additional emission channels are allowed for $\ell=1-5$ (with $n=1$) and $\ell=46-50$ (with $n=2$), as marked by filled blue dots, thus giving rise to a total of $10$ additional emission channels through which the array can radiate, and are represented as arrows outgoing along direction $\theta_{n\ell}$ in Fig.~\ref{fig1}(d). In orange, we present an additional case where we choose $a/\lambda=0.25$ such that no constructive interference condition can be met, for any value $n$, when $\ell=0$ (i.e., Eq.~\eqref{eq:SP0condition} has no real solutions and thus no traditional SP emission can take place). Nevertheless, we find $16$ allowed GSP emission channels (filled orange dots), for $n=3$ and $\ell=13-38$, whose emission direction spans the entire $\phi=0$ plane ($-90\mydeg\leq\theta_{3,\ell}\leq 90\mydeg$), as visible in Fig.~\ref{fig1}(e). 

\begin{figure*}[htbp]
    \centering
    \includegraphics[width=\linewidth]{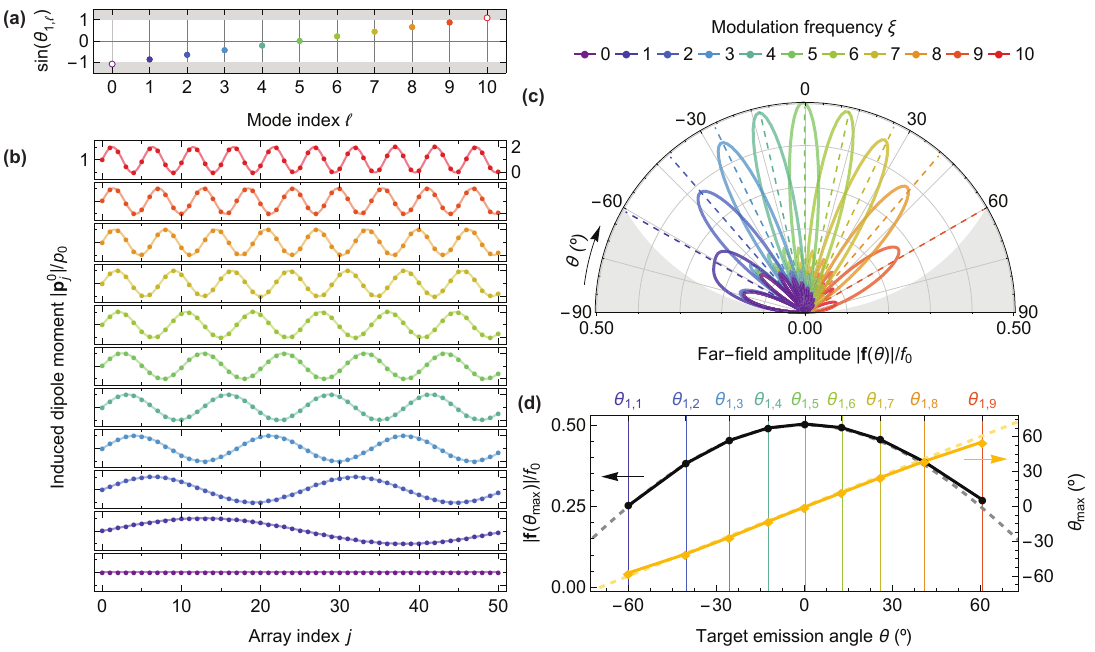}
    \caption{\textbf{Cathodoluminescence steering with non-uniform arrays.} \textbf{(a)} GSP condition in Eq.~\eqref{eq:SPcondition} for $n=1$ and $\ell=0-10$ in an array with $N=51$ elements and period $a/\lambda = 0.090$, excited by an electron beam with velocity $v=0.1\,c$ ($\approx 2.6$~keV). Modes $\ell=1-9$ lie within the region $|\sin \theta_{1,\ell}|\leq 1$ and are marked by filled circles, whereas modes $\ell=0$ and $\ell=10$ lie outside that region and are marked with empty circles. Each mode is targeted by the same-color induced dipole moment distribution in (b) that selects the mode $\ell=\xi$. \textbf{(b)} Induced dipole moment distributions following Eq.~\eqref{eq:pjharmonic} (with oscillation amplitude $A=1$) for the color-coded values of $\xi$ indicated by the legend, as a function of array element index $j$. Each distribution of induced dipole moment $|\pb_j^0|/p_0$ oscillates around $1$ between $0$ and $2$ with a frequency $\xi$. \textbf{(c)} Angle-resolved cathodoluminescence far-field emission amplitude $|\fb(\theta)|$ polar plot, normalized to $f_0=Nk^2p_0$, as a function of the $\theta$ angle in the $\phi=0$ plane, for an electron passing above an array whose induced dipole moments are given by the same-color curve in (b) and polarized along $x$ ($\pb_j^0 \parallel \xx$). The colored dashed lines mark the position of the target angles $\theta_{1,\xi}$ for $\xi=1-9$ and the shaded gray area represents the condition $|\fb(\theta)|/f_0 > \cos(\theta)$. \textbf{(d)} Peak emission angle $\theta_{\rm max}$ (right axis) and corresponding peak far-field amplitude $|\fb(\theta_{\rm max})|/f_0$ (left axis) as a function of the target emission frequency $\theta$ for $\xi=1-9$ (see top axis). The dashed gray curve represents the function $|\fb(\theta)|/f_0=\cos(\theta)/2$ and the yellow dashed line corresponds to the condition $\theta_{\rm max}=\theta$. }
    \label{fig2}
\end{figure*}

\subsection{Steering of CL using non-uniform arrays}

Remarkably, the GSP condition in Eq.~\eqref{eq:SPcondition} is a purely geometrical property of the array and is independent of the polarization of the individual dipoles $\pb_j^0$, which affects the emission properties along a given channel, but not the existence of the channel itself. Parameters $v/c$, $a/\lambda$, and $N$ fully control the emergence and properties of the GSP channels, as long as the array is periodic and its elements are dipolar in nature. In turn, access to the $\ell$th emission channel (for any $n$) is controlled by the corresponding $\ell$th GSP component
\begin{align}
\tilde{\pb}_{\ell}=\frac{1}{N}\sum_{j} {\pb}_j^0 \ee^{-2\pi \ii j \ell/N},
\label{eq:pell}
\end{align}
which depends on the distribution of the phase-corrected induced dipole moments $\pb_j^0$ across the array, governed by the physical properties of the array elements and all intra-array interactions. This means that, to target a specific emission channel denoted by an $\{n,\ell\}$ pair, the induced dipole moment distribution in the array must be engineered to yield a strong Fourier component along harmonic mode $\ell$.

We discuss in the next section different ways in which, in practice, one may design the array to achieve a given dipole moment distribution $\pb_j$ in an experimental setup. For now, we assume that we have full control over such distribution. One simple option to achieve the targeting of a specific individual $\ell$ mode is to design the array such that the induced dipole distribution follows a harmonic dependence, such as
\begin{align}
  \frac{|\pb_j(\xi)|}{p_0} = 1+A\sin\left(\frac{2\pi \xi j}{N}\right) ,
  \label{eq:pjharmonic}
\end{align}
whose Fourier transform directly yields $\tilde{p}_l/p_0 = \delta_{\ell,0}+A(\delta_{\ell,\xi}-\delta_{\ell,N-\xi})/2\ii$, with $\tilde{p}_l = |\tilde{\pb}_l|$. Here, $\xi$ is an integer parameter that represents the frequency of modulation of the dipole moment distribution, which oscillates around the baseline value $p_0$ with some amplitude $|A|\leq 1$. By choosing the value of $\xi$, we are able to access the GSP modes with $\ell=\xi$ and $\ell=N-\xi$, and hence the system can radiate along angles $\theta_{n,\xi}$ and $\theta_{n,N-\xi}$ (if the corresponding channel is available for the specific parameters of the array and electron). Importantly, the nonzero baseline of the distribution additionally enforces that the $\ell=0$ mode is available for any value of $\xi$, and therefore, traditional SP emission is always present (as long as it is kinematically allowed).

In what follows, we apply the GSP formalism to steer CL for an electron beam passing near a periodic array. We consider a periodic array with $N=51$ scatterers and period $a/\lambda=0.090$ at some emission wavelength $\lambda$, and an electron moving with velocity is $v=0.1\,c$, whose total available GSP modes correspond to the choice $n=1$ and $\ell=1-9$, as shown in Fig.~\ref{fig2}. It is important to note that we choose $\lambda/a$ to purposely exclude the $\ell=0$ channel and therefore remove the persistent signature of traditional SP. To target each mode, we adopt an induced dipole moment distribution with the form of Eq.~\eqref{eq:pjharmonic}, with the parameter $\xi$ varying from 0 to 10, giving rise to the distributions shown in Fig.~\ref{fig2}(b). While Eq.~\eqref{eq:pjharmonic} prescribes solely the amplitude of each dipole moment, we arbitrarily take all dipoles to be polarized along $x$ (that is, $\pb_j = |\pb_j| \xx$). We discuss the implications of the dipole moment orientation below.

For each dipole moment distribution, we compute the corresponding far-field emission $\fb(\theta)$ as described, normalized to the reference value $f_0=Nk^2p_0$, as a function of the polar angle $\theta$ and in the $\phi=0$ plane, as shown in Fig.~\ref{fig2}(c). We present only emission along the positive $z$ direction ($-90\mydeg \leq\theta\leq 90\mydeg$), since emission along the negative $z$ axis follows symmetrically. The curves for $\xi=0$ and $\xi=10$ correspond to cases with no viable GSP emission channel, and thus, there is no pronounced emission at a specific angle. Nevertheless, small residual lobes appear because the array has a finite number of elements $N$, which prevents total destructive interference. In contrast, for $\xi=1-9$, we observe clear lobes indicating a resonance in the angular emission at well-defined and nearly equally spaced angles ranging between $\sim-60\mydeg$ and $\sim 60\mydeg$, matching very closely the targeted angles $\theta_{1,\xi}$ marked by dashed lines. This behavior is clearer in Fig.~\ref{fig2}(d), where we plot the peak far-field amplitude (left axis) and the observed peak emission angle (right axis) as a function of the targeted angle $\theta_{1,\xi}$ for each $\xi=1-9$. The agreement between target and observed angles is very remarkably accurate, with small deviations (most evident at the smallest and largest targeted angles) stemming from the finite number of array elements (see discussion below). Furthermore, we observe that, in all cases, the peak amplitude follows very closely a $|\cos\theta|$ envelope (dashed gray line) with respect to the theoretical maximum $f_0/2$ (where the $1/2$ comes from the Fourier transform of Eq.~\eqref{eq:pjharmonic}). This envelope, arising from the $|(1-\rr\otimes\rr)\cdot\xx|$ term in Eq.~\eqref{eq:SPf} for $x$-oriented dipoles, promotes emission into the normal direction ($\theta=0\mydeg$) while suppressing emission into the grazing directions ($\theta=\pm90\mydeg$) in the $\phi=0$ plane, similarly to the behavior of a single $x$-polarized point dipole.

The number of array elements $N$ emerges as a crucial parameter determining both the CL steering range and angular resolution. The larger $N$ is, the more kinetically-allowed GSP modes exist, and they sample more compactly the kinetically allowed region $|\sin\theta_{n\ell}|\leq 1$. This results in smaller gaps between consecutive emission lobes while increasing the steering range, as evident in Fig.~\ref{figS1} in the Appendix, which presents an analogous analysis to that of Fig.~\ref{fig2} for an array with $N=101$ elements (instead of $N=51$). As shown in Fig.~\ref{figS1}(a), the larger array supports more viable GSP modes, resulting in a consequent reduction in inter-mode spacing. In Fig.~\ref{figS1}(b), we plot the far-field emission for $x$-oriented dipoles whose induced dipole moment follows Eq.~\eqref{eq:pjharmonic}, showing that the number of discrete steering lobes increases, and the lobes become narrower, more closely-spaced, and match more accurately the target angles, while the far-field amplitude remains bounded by the $\cos\theta$ envelope (see also Fig.~\ref{figS1}(d)). The steering range for $N=101$ spans approximately $-70\mydeg<\theta<70\mydeg$, which, however, represents only a mild improvement when compared with $N=51$ in Fig.~\ref{fig2}. This small increase when approximately doubling $N$ is explained by the large slope of the $\arcsin(x)$ function near $x=\pm 1$, which needs increasingly larger $N$ to reach steering along more grazing angles (closer to $\pm90\mydeg$). In Fig.~\ref{figS1}(c), we show similar results as in \ref{figS1}(b) but for dipoles oriented along $z$. While the available steering angles remain the same (they are independent of the specific distribution $\pb_j$), the corresponding CL far-field amplitude at the plane $\phi=0$ follows now an envelope $|(1-\rr\otimes\rr)\cdot\zz|=|\sin\theta|$ (see also Fig.~\ref{figS1}(e)) that suppresses emission into the normal direction and benefits emission along grazing angles. However, the inherent challenge in opening channels for emission at grazing angles discussed above makes $z$-oriented dipoles less suitable for wide-range CL steering. Instead, array elements designed to exhibit a dominant dipolar response along the $x$ direction should be more efficient for such a task.

In Fig.~\ref{figS2} in the Appendix, we highlight two additional properties of the GSP emission. Firstly, Fig.~\ref{figS2}(a) shows that the modulation amplitude $A$ in Eq.~\eqref{eq:pjharmonic} directly controls the far-field peak amplitude, with a larger modulation leading to stronger steering (i.e., larger peak amplitude), but not altering the emission angle. Secondly, Fig.~\ref{figS2}(b) addresses the case where the induced dipole moment distribution is characterized by a superposition of several $\xi$ values, which, by the linearity of the Fourier transform, results in the simultaneous emission of CL along each mode $\xi$ in the superposition. Then, the associated far-field amplitude along each mode becomes, in that case, proportional to the corresponding superposition coefficient. Both of these properties suggest that highly complex emission patterns can be achieved by engineering the induced dipole moment distribution in suitable manners that incorporate different harmonic frequencies $\xi$ and amplitudes $A$, and possibly combine dipoles polarized along the $x$ and $z$ directions, thereby enabling tailored angular responses without modifying the underlying array geometry.

Finally, although the results above were derived for an array of point-like (0D) scatterers, they extend directly to one-dimensional dipole lines that run indefinitely in the $y$ direction. As detailed in Methods and the Appendix, the induced dipole on line $j$ can be expanded as $\pb_{j,q}$, labeled by the longitudinal wave vector $q$ arising from the translational invariance of the system along the $y$ direction. This dipole component is excited by the same component of the external electric field given by Eq.~\eqref{eq:Eextq}, where the angles $(\theta,\phi)$ of the far-field CL are identical to the point-dipole result once the component with $q = k\sin\theta\sin\phi$ is selected. Consequently, all design rules in this section are readily adapted to 1D geometries (such as wires or ribbons), as we explore in the next section. 

\subsection{Engineering non-uniform induced dipole distributions}

Up to this point, we have focused on describing the GSP effect and its applications, starting from a finite array with an induced dipole moment distribution presumed to have been previously engineered. Now, we turn our attention to the ways in which one can engineer such arrays in practice, and, in particular, we study physical setups that can be actively tuned to achieve a dynamically induced dipole moment distribution.

The dipole moment $\pb_j=\pb_j^0 \ee^{\ii \omega j a/v}$ induced in a given element of the array, given by Eq.~\eqref{eq:pj}, is primarily governed by its polarizability $\alpha_j$, which captures how the element responds to an external field depending on its geometry, material, and electromagnetic environment. Consequently, realizing a prescribed distribution set $\{|\pb_j|\}$ reduces to inverse-designing a compatible set $\{\alpha_j\}$ that self-consistently generates it. Although one could employ numerical or machine-learning frameworks for this step, we present in the Methods section an analytical simplification that yields closed-form polarizability prescriptions under certain conditions. The resulting design is then implemented in the array by choosing the physical properties of each scatterer, either \textit{passively} --by modifying the element geometry (length, width, thickness), material composition or doping (hence its dispersion), and the surrounding electromagnetic environment (substrate permittivity, spacer thickness, cavity backing)-- or \textit{actively} --via dynamic tuning such as electrostatic gating, optical/thermal pumping, or voltage biasing.

Here, we focus mainly on active tuning, as we envision that it allows for on-demand control of the array properties that, combined with the results from the previous section, could be applied to realize active steering of CL radiation, among many other technologically-appealing possibilities. Furthermore, we focus on systems composed of 2D or quasi-2D materials naturally exhibiting exclusively in-plane polarization, which simplifies the inverse-design process (see Methods) and is more suitable for tunable emission of radiation at near-normal angles, as discussed above. Specifically, we consider arrays of VO$_2$ disks whose permittivity can be adjusted by controlled fluence exposure, and arrays of graphene ribbons whose Fermi level can be tuned by electrostatic gating (see Methods). In both types of systems, we preserve the array number of elements $N=51$, electron velocity $v=0.1\,c$ ($\approx 2.6$~keV), and period-to-wavelength ratio $a/\lambda=0.090$ chosen in Fig.~\ref{fig2} and, as such, the allowed GSP emission channels ($n=1$, $\ell=1-9$, see Fig.~\ref{fig2}(a)) remain unchanged. As examples, we reverse-design the fluence dosage per disk and the Fermi level per ribbon necessary to imprint in the array an induced dipole moment distribution given by Eq.~\eqref{eq:pjharmonic} with $\xi=3$ and $\xi=7$, yielding a target emission along angles $\theta_{1,3}=-25.5\mydeg$ and $\theta_{1,7}=26.0\mydeg$.

\begin{figure*}[htbp]
    \centering
    \includegraphics[width=\linewidth]{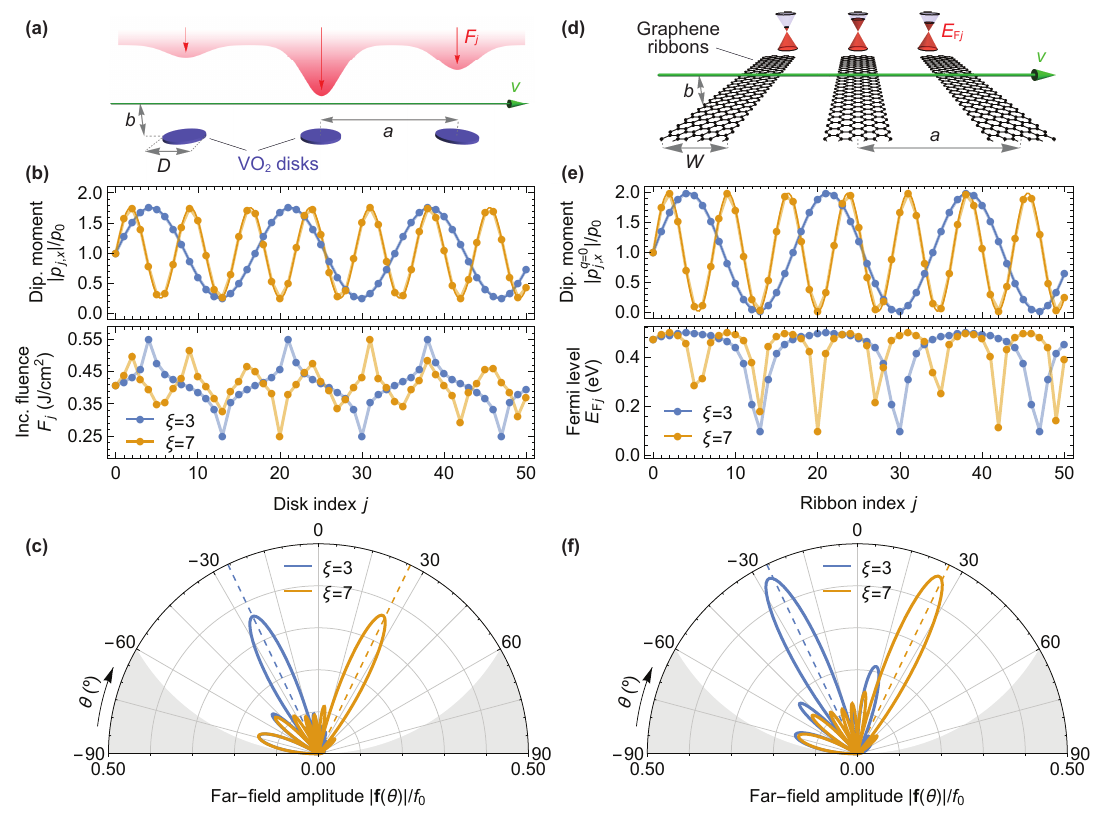}
    \caption{\textbf{Active tuning of CL emission.} \textbf{(a)} Scheme of an array of VO$_2$ disks with diameter $D=250~$nm and thickness $t=2$~nm, separated from their nearest neighbors by a center-to-center distance $a=450~$nm, with an electron passing parallel to the array at a distance $b=10$~nm and with velocity $v=0.1\,c$. The array is illuminated by a pump beam (in red) that is spatially engineered to deliver a fluence $F_j$ at the $j$th array element. \textbf{(b)} Induced dipole moment at a wavelength $\lambda=2\pi c/\omega=5.0~\mathrm{\mu m}$ (top) along different array elements, following Eq.~\eqref{eq:pjharmonic} with $\xi$ values as indicated by the labels, $A=1$, and $p_0=5.4\times 10^{-3} (eD/\omega)$, as a function of element $j$, for the color-coordinated fluence distributions (bottom), ranging between $F_{\rm min}=0.25~\mathrm{J/cm^2}$ and $F_{\rm max}=0.55~\mathrm{J/cm^2}$. \textbf{(c)} Far-field emission distribution for the same-color array distributions in (b), with $f_0=2.7\times 10^{-2}(e/D\omega)$. \textbf{(d--f)} Same as (a--c), but for an array of graphene ribbons with width $W=100$~nm whose $j$th element's Fermi level is set to $E_{{\rm F}j}$ (as depicted by the Dirac cones) between $E_{\rm F,min}=0.1$~eV and $E_{\rm F,max}=0.5$~eV, with a period $a=615~\mathrm{nm}$ and evaluated at a wavelength $\lambda=6.8~\mathrm{\mu m}$. In (e), the dipole moment corresponds to the $q=0$ component, with $p_0=4.48(eW/\omega)$. In (f), we have $f_0=1.95/(e/W\omega)$.}
    \label{fig3}
\end{figure*}

\subsubsection{VO$_2$ disks}

Vanadium dioxide (VO$_2$) is a phase-changing material that undergoes a rapid, reversible insulator--metal transition (IMT) at a modest temperature ($\sim 340$~K) \cite{CGL20,DPQ08}. Such transition is accompanied by an abrupt change in optical and electrical properties, with VO$_2$ exhibiting markedly different dielectric functions at the insulating and metallic phases \cite{VBB68,QBC07}, as shown in Fig. \ref{figS3}(a) in the Appendix. Nevertheless, intermediate states in between the two phases can be accessed by controlling the temperature of the material, and are typically parameterized by a metallic volume fraction $f_{\rm m}$ ranging between $0$ and $1$ (see Methods). As the material cools down, it reverts slowly to the insulator phase, with a characteristic time scale that depends on how quickly the structure releases heat, ranging from tens to hundreds of nanoseconds in thin films on substrates \cite{WGB13,LKR17} to microseconds in suspended or weakly coupled nanostructures \cite{MBW16}. VO$_2$ further counts with mature thin-film growth and integration on common substrates \cite{MAJ14}, and reliable nanofabrication with reversible optical/electrical/thermal switching demonstrated from ultrafast to microsecond regimes \cite{JFT06,PKE11,WGB13,DPQ08,MBW16,CGL20}, which make it a suitable candidate for active switching applications.

Figure~\ref{fig3}(a) illustrates an array of VO$_2$ disks with a diameter of $D=250$~nm, thickness of $t=2$~nm \cite{MAJ14}, and period $a=450$~nm. The structure is illuminated by an incident pump optical pulse whose wavefront is engineered to deliver specific energy doses to different disks, thereby inducing spatially varying local temperatures. As a result, each disk experiences a distinct optical excitation fluence $F_j$, leading to a controlled modification of its permittivity through the thermally driven IMT in VO$_2$. The variation in the polarizability of such disks with incident fluence at the target probe wavelength $\lambda=5.0~\mathrm{\mu m}$ (chosen to yield a large contrast between the insulating and metallic phases of the VO$_2$) is shown in Figure~\ref{figS4}(a) in the Appendix (see also Methods) for a pump laser with wavelength $\lambda_{\rm pump}=632~\mathrm{nm}$. In practice, such modulation of the excitation can be customized using a spatial light modulator, controlled interference of multiple beams, or active metamaterial masks, provided that the required spatial pattern does not contain features below the optical diffraction limit of the pumping wavelength (i.e., if $a \gtrsim  \lambda_{\rm pump}$).

Under illumination by an electron beam with impact parameter $b=10$~nm, as also shown in Fig.~\ref{fig3}(a), the interference pattern of the emitted CL depends on the fluence dose delivered to each disk. We note that the electron traverses the structure in a timescale $Na/v$ of the order of a few picoseconds, during which the permittivity of the disks can be regarded as approximately constant. In Fig.~\ref{fig3}(b) (bottom panel), we show two distinct fluence-delivery patterns that were inverse-designed as described in the Methods to generate, using Eqs.~\eqref{eq:pj} and \eqref{eq:Eext}, the dipole moment distributions in the color-coordinated curves in the top panel. As warranted, the reconstructed induced dipole moment distribution successfully reproduces Eq.~\eqref{eq:pjharmonic} with $\xi=3$ (blue) and $\xi=7$ (orange). Although dipole-dipole interactions were considered for the reconstruction algorithm (see Methods), the relatively small dipole moment in each disk and the large separation between consecutive disks made them negligible.

Using Eq.~\eqref{eq:SPf}, we finally calculate the resulting far-field emission patterns, displayed in Fig.~\ref{fig3}(c), confirming that the emitted radiation is strongly concentrated around the desired target angles $\theta_{1,3}$ and $\theta_{1,7}$, demonstrating excellent agreement between the designed and realized responses. The other available emission angles for the considered geometrical parameters ($\xi=1-9$, as in Fig.~\ref{fig2}) can be targeted analogously. We note that the emission amplitude falls short of its theoretical maximum value $f_0/2$ due to a finite background dipole moment arising from the weakly polarizable insulating phase. As a result, although the metallic phase exhibits a substantially larger polarizability, the modulation depth $A$ of the dipole amplitude in Eq.~\eqref{eq:pjharmonic} cannot reach unity [$A\approx 0.75$ for the example in Fig.~\ref{fig3}(b)].

\subsubsection{Graphene ribbons}

Graphene nanoribbons constitute a highly versatile alternative platform for implementing the type of spatially controlled optical modulation of CL. Their optical and electronic properties can be tuned continuously through electrostatic gating, chemical doping, or strain, offering direct control over the charge carrier density and consequently the induced dipole moment \cite{paper235,GPN12}. In contrast to confined structures such as nanodisks, ribbons support extended plasmonic modes with well-defined momentum along their length, which can be exploited to engineer collective responses \cite{paper257,paper364}. Furthermore, graphene ribbons are particularly well suited for experimental realization, since they can be fabricated with high precision using established lithographic and transfer techniques \cite{JZW09,CRF10}, integrated into a variety of substrates \cite{SKM11,MLJ21}, and interfaced readily with local gate electrodes \cite{ZQB23,KJB16}. With suitable device design \cite{STG24}, the Fermi level of each ribbon can be individually adjusted, enabling spatially resolved and actively reconfigurable modulation patterns. These attributes make graphene nanoribbons an attractive and practical platform for dynamic control of light–matter interactions.

In Fig.~\ref{fig3}(d), we present an example of an array of graphene nanoribbons with a width of $W=100$~nm and a period of $a=615~\mathrm{nm}$. The ribbons in this array are assumed to be individually tunable, allowing independent control of their Fermi levels. We target an optical wavelength of $\lambda=6.8~\mathrm{\mu m}$ ($\approx 0.18$~eV) corresponding to the resonance wavelength for a $0.5$-eV-doped graphene ribbon, as shown in Fig.~\ref{figS4}(b) in the Appendix.  Figure~\ref{fig3}(e) (bottom) illustrates two representative Fermi-level profiles for an array comprising $N = 51$ ribbons, ranging between $0.1$ and $0.5$~eV, which are designed to produce the dipole-moment distributions shown in the top panel following Eq.~\eqref{eq:pjharmonic}, with $\xi = 3$ (blue) and $\xi = 7$ (orange). We note that such dipole moment corresponds to the $q=0$ component (see Appendix) that controls the GSP emission along the $\phi=0$ plane, as discussed above. As in the disks geometry, dipole-dipole interactions were included in the inverse design algorithm, but turned out to be negligible for the chosen geometric parameters. However, in contrast to the disks, here the modulation depth $A$ can approach unity, as the polarizability of each graphene ribbon is strongly dependent on its doping level and nearly vanishes when off-resonance.  Figure~\ref{fig3}(f) displays the corresponding far-field emission patterns, demonstrating that the radiated intensity is concentrated around the desired target angles $\theta_{1,3}$ and $\theta_{1,7}$, in excellent agreement with the designed emission directions. 

\subsubsection{Alternative passive methods}

In the Appendix, we present for completeness two examples of passive engineering, which involve adjusting the length of small silver nanorods and the width of thin gold ribbons, whose polarizability can be controlled through their dimensions as shown in Fig.~\ref{figS4} in the Appendix. As above, we consider arrays with $N=51$ elements, a ratio $a/\lambda=0.090$, and $v=0.1\,c$. While these structures are not actively tunable, they have the advantage of being conceptually simpler and potentially easier to realize. In Fig.~\ref{figS5}, we show the inverse design of the physical dimensions of each scatterer, leading again to a highly directional emission of light that follows closely the target angles for each distribution. Interestingly, the large dipole moment achieved by the noble metals considered in these examples renders the dipole-dipole interactions non-negligible, but the procedure explained in the Methods section is able to account for those successfully, as proven by the reconstructed dipole moment distribution.

Observing the designed length profile of the nanorods, one notes that reproducing the theoretically prescribed geometry could demand high fabrication precision. This is because even a small deviation in length can lead to a significant change in the rod polarizability (see Fig.~\ref{figS4} in the Appendix), thereby affecting the intended dipole distribution. While this effect is particularly pronounced for silver nanorods, it also applies to gold and graphene nanoribbons, where the response relies on the strong dipolar resonances supported by these structures. Nevertheless, this challenge can be mitigated by inverse-designing the polarizability in a regime where its dependence on the control parameter (length, width, or Fermi level) is smoother (i.e., slightly off-resonant) or by targeting broader lower-quality-factor resonances. Such approaches should relax fabrication tolerances at the cost of a reduced induced dipole amplitude and, consequently, lower light-emission efficiency, which can however be compensated by adjusting the intensity or impact parameter of the electron beam.

\section{Discussion} 

We have introduced a generalized framework for SP radiation in finite and non-uniform arrays, extending the classical concept to structures with spatially varying dipole moments. This GSP condition enables emission into non-traditional angles and spectral channels, whose accessibility is determined by the engineered amplitude and phase distribution of the dipoles. By prescribing sinusoidal modulations of the dipole moment across the array, we have demonstrated the ability to steer CL emission actively and predictably, thereby establishing a versatile approach to programmable free-electron light sources. Nevertheless, our concept and methods can be straightforwardly generalized to other types of modulation profiles.

Two representative active implementations were presented: arrays of VO$_2$ nanodisks and graphene nanoribbons. In VO$_2$, patterned optical excitation enables spatial modulation of the local permittivity through thermally driven insulator--metal transitions, while in graphene, the dipole response can be tuned continuously through electrostatic gating. Both systems achieve targeted angular emission in agreement with the GSP theory. We find that graphene offers nearly complete modulation depth owing to its highly controllable polarizability. Collectively, these results highlight the remarkable degree of control attainable with actively tunable materials whose polarizability can be controlled \textit{in situ}. By adjusting the optical properties of the array elements in real time, one can switch between different excitation profiles and, consequently, between distinct dipole-moment distributions for tailoring free-electron–driven light emission. By increasing the number of array elements, the degrees of freedom to achieve more complex emission patterns is equally increased. Some additional degree of tunability can still be achieved by adjusting the velocity $v$ and impact parameter $b$ of the electron beam. Beyond the examples presented in this paper, a broad range of alternative materials and mechanisms may be harnessed to realize GSP modulation, including thermal or electrostatic tuning of transition-metal dichalcogenides (TMDs), phase-change chalcogenides, or photo-doped two-dimensional materials such as graphene and MoS$_2$. Alternatively, instead of tuning the emitters themselves, one could control the surrounding environment using active substrates that can be spatially modulated, thereby enabling additional degrees of reconfigurability.

The formalism developed in this work is general and can be extended beyond the electrostatic approximation used for analytical traceability of the scatterers. While we have modeled each scatterer as a point (or line) dipole exhibiting a quasistatic response, the main features of the GSP condition and the resulting directional emission should remain valid when retardation and finite-size effects are taken into account, albeit the precise resonance conditions may shift slightly. Moreover, although the present work focuses on effectively one-dimensional arrays consistent with the geometry defined by a single electron trajectory, the theoretical framework is readily adaptable to more complex excitation schemes, including multi-beam or two-dimensional arrangements, which could enable richer control over both azimuthal and polar emission angles. 

These results highlight a powerful strategy for bridging near-field electron excitation with metasurface design. By tailoring the local response of individual elements, one can dynamically reconfigure the collective radiation pattern without altering the underlying geometry. Such control opens new possibilities for reconfigurable free-electron light sources, adaptive beam steering, spectral shaping, and on-demand generation of structured or directional emission. As an extension of this concept, we envision the control of 2D arrays, such that the electron can produce emission over controlled directions in the 2D angular hemisphere. In the longer term, this approach may enable electron-driven nanophotonic devices for active holography, ultrafast sensing, and programmable quantum light generation.

\section{Materials and methods} 

\subsection{Field generated by a moving electron}

We consider an electron moving along a trajectory defined by $y=0$ and $z=b$, with constant velocity $v$ along $x$ (see Fig~\ref{fig1}(b)). The electric field generated by such an electron at frequency $\omega$ and position $\rb=x \xx+\Rb$ is given by \cite{paper149}
\begin{align}
 \Eb^{\rm ext}(\rb) = \frac{2e\omega}{v^2\gamma}\left[\frac{\ii}{\gamma} K_0\left(\frac{\omega R}{v \gamma}\right)\xx - K_1\left(\frac{\omega R}{v \gamma}\right)\RR\right] \ee^{\ii\omega x/v},
 \label{eq:Eext}
\end{align}
where the spatial coordinates are defined as $\Rb=y \yy+(z-b)\zz$, $R=|\Rb|$, $\RR=\Rb/R$, and the Lorentz factor $\gamma=1/\sqrt{1-v^2/c^2}$.

To describe translationally-invariant systems along $y$ (e.g., ribbons) illuminated by the electron, it is convenient to expand the electric field as $\Eb^{\rm ext}(\rb)=(2\pi)^{-1} \int {\rm d}q\, \Eb^{\rm ext}_q(x,z)\ee^{\ii q y}$, where $q$ is the wave vector along $y$, and
\begin{align}
  \Eb^{\rm ext}_q(x,z) = \frac{2 \pi \ii e}{v \kappa_z} \left( \frac{\omega}{v \gamma^2}, q, \ii \kappa_z  \right)\ee^{-\kappa_z |z-b|} \ee^{\ii \omega  x/v},  
  \label{eq:Eextq}
\end{align}
and we introduce $\kappa_z^2= \omega^2/v^2 \gamma^2+q^2$.

\subsection{Generalized Smith-Purcell effect with 1D scatterers}

We extend the GSP condition for point dipoles to an array of line dipoles placed periodically along $x$, with a period $a$, infinitely extended along $y$, and located in the plane $z=0$. In practice, each line can represent a ribbon or cylinder that is sufficiently narrow and well separated from its neighbors such that, in the far field, it can be approximated by a line with negligible lateral extent. As above, we take the electron moving parallel to $x$ with velocity $\vb=v \xx$ along a trajectory defined by $y=0$ and $z=b$.

Unlike the point dipole structure, now the line $j$ exhibits a dipole moment density $\Pb_{j,y}$ (per unit length along $y$), which depends on the coordinate $y$. Following the same procedure as above, we anticipate that the finite electron velocity leads to a phase difference to be imprinted on the dipole moment induced in each line, so we write $\Pb_{j,y}=\Pb_{j,y}^0 \ee^{\ii \omega j a/v}$. Then, the differential far-field intensity generated by a dipole element $dy$ at $y$ can be written analogously to Eq.~\eqref{eq:SPf} as
\begin{align}
  {\rm d}\fb(\rr) = k^2 (1-\rr\otimes\rr) \cdot \sum_j \Pb_{j,y}^0 \ee^{\ii k j a/\beta} \ee^{-\ii k (\rb_j \cdot \rr)} {\rm d}y,
  \label{eq:dfline}
\end{align}
where $\rb_j=(j a, y, 0)$, and the total far field follows as $\fb(\rr)= \int {\rm d}y\, [{\rm d}\fb(\rr)/{\rm d}y]$.

To simplify the equation above, we note that the translational symmetry of the problem allows us to write, in general, the $y$-dependence of the dipole moment of line $j$ in terms of a wave vector $q$ as
\begin{align}
  \Pb_{j,y} = \frac{1}{2\pi} \int {\rm d}q\, \Pb_{j,q} \ee^{\ii q y},
  \label{eq:pjy}
\end{align}
whose expansion components $\Pb_{j,q}$ represent the eigenmodes of each line element, parameterized by $q$, and whose profiles can be determined depending on its material and geometric parameters (such as the permittivity/conductivity and dimensionality). When arranged in an array with period $a$, the self-consistent induced dipole moment density at each ribbon and for momentum component $q$ follows from the self-consistent equation $\Pb_{j,q}=\alpha_{j,q}\cdot[\Eb^{\rm ext}_q(ja,0)+\sum_{i\neq j}\mathcal{G}_{q,ji}\cdot \Pb_{i,q}]$ [i.e., analogous to Eq.~\eqref{eq:pj} for point dipoles], where the $q$-component electron field $\Eb^{\rm ext}_q$ is shown above, and the corresponding ribbon polarizability $\alpha_{j,q}$ and line dipole-line dipole interaction Green tensor $\mathcal{G}_{q,ji}$ are discussed in the Appendix. 

Replacing Eq.~\eqref{eq:pjy} into Eq.~\eqref{eq:dfline} and performing the integration over $y$, we obtain a term with the form $\int {\rm d}y\, \ee^{\ii q y}  \ee^{-\ii k y(\yy \cdot \rr)} = 2\pi\delta(q-k_y)$, where we define $k_y=k(\yy \cdot \rr)=k \sin\theta\sin\phi$. At this point, the Dirac delta function resolves the $q$-integral, yielding the final expression
\begin{align}
  \fb(\rr) = k^2 (1-\rr\otimes\rr) \cdot \sum_j \Pb_{j,k_y}^0 \ee^{\ii k j a/\beta} \ee^{-\ii k a j(\xx \cdot \rr)}.
  \label{eq:ff:line}
\end{align}
Equation~\eqref{eq:ff:line} is remarkably similar to Eq.~\eqref{eq:SPf} (for an array of point dipoles) with the exception that, for a given outgoing direction $\rr$, we must take only the $q=k (\yy\cdot\rr)$ component out of the full momentum distribution of $\Pb_{j,y}$ in Eq.~\eqref{eq:pjy}. In particular, the SP pattern remains exactly the same as before in the $\phi=0$ plane when only the $q=0$ components are considered.

\subsection{Inverse design of the polarizability}

We aim to map a given phase-corrected induced dipole moment distribution $\pb_j^0=\pb_j \ee^{-\ii \omega ja/v}=p_{j,x}^{0}\xx+p_{j,z}^{0}\zz$ onto a polarizability tensor distribution, where each element has components $\alpha_j^{\nu\mu}$ ($\nu,\mu=x,y,z$). For simplicity, we restrict this analysis to diagonal polarizability tensors obeying the property $\alpha_j^{\nu\mu}=\alpha_j^{\nu\nu}\delta_{\mu\nu}$. 

In general, we cannot prescribe both components of the dipole moment, since they are not independent, but must rather preserve the self-consistency of the coupled-dipole formalism with respect to the external field [in this case, the electron beam, whose field is given by Eq.~\eqref{eq:Eext}], and the dipole-dipole interaction between the different elements of the array. Furthermore, in general, one may aim to achieve a specific dipole moment profile up to a global scaling constant [eg., $p_0$ in Eq.~\eqref{eq:pjharmonic}] that must be determined self-consistently. From these two considerations, we can write $p^0_{j,\nu}=p_0 b_{j,\nu}$ and assume that all elements $b_{j,\nu}$ are prescribed for one specific component $\nu$. From here, we aim to reconstruct some physically compatible polarizability components $\alpha^{\nu\nu}_j$, together with $p_0$ and the remaining induced dipole components.

Introducing $\Eb_0^{\rm ext}=\Eb^{\rm ext}(\rb_j)\ee^{-\ii \omega j a/v}$ (which is uniform across the array) and the notation defined above, we can use Eq.~\eqref{eq:pj} to obtain the relation
\begin{align}
\alpha^{\nu\nu}_j=\frac{p_0 b_{j,\nu}}{E_{0,\nu}^{\rm ext}+p_0 \sum_{i}\sum_{\mu} \mathcal{G}^{\nu\mu}_{ji} b_{i,\mu}}.
\label{eq:alphasol}
\end{align}
While this equation precisely relates the polarizability and self-consistent dipole moment across an array [it is equivalent to Eq.~\eqref{eq:pj}], it is not general enough to reconstruct unequivocally $\{b_{j,\nu}\}$ into $\{\alpha_j^{\nu\nu}\}$ because it yields separate prescriptions for $\alpha_j^{xx}$ and $\alpha_j^{zz}$ that must be mutually compatible with the material and geometry chosen and the external field components. Therefore, Eq.~\eqref{eq:alphasol} (for $\nu=x,z$) must be supplied by additional information regarding the symmetry of the polarizability tensor (e.g., $\alpha_j^{xx}=\alpha_j^{zz}$ if it is isotropic) to form a nonlinear system of $3N$ equations that can be numerically solved to yield both components of the polarizability tensors and the non-prescribed component of the induced dipole moment. Finally, $p_0$ can be set by imposing an additional condition: for example, the array element where the induced dipole is maximum (according to the prescription) must exhibit the maximum designed polarizability allowed by our physical system.

While the process described above is in general computationally heavy for more than a few array elements, there are two scenarios in which Eq.~\eqref{eq:alphasol} can be simplified and employed directly:
\begin{itemize}
\item When the array elements are sufficiently far apart, their dipole-dipole interactions become negligible and Eq.~\eqref{eq:alphasol} simplifies to $\alpha^{\nu\nu}_j=p_0 b_{j,\nu}/E_{0,\nu}^{\rm ext}$, which decouples the $x$- and $z$-components. Ensuring self-consistency becomes trivial: for an isotropic tensor, for example, where the coefficients $b_{j,x}$ are prescribed, it suffices to set $b_{j,z}=b_{j,x}(E^{\rm ext}_{0,z}/E^{\rm ext}_{0,x})$ and $p_0=\alpha_{\rm max}E_{0,x}^{\rm ext}/b_{\rm max}$, with $b_{\rm max}$ denoting the maximum value in the prescribed set $\{b_{j,x}\}$, and $\alpha_{\rm max}$ the maximum polarizability of the array elements.

\item If the scatterers are highly anisotropic, such that one component $\nu$ of the polarizability dominates, while the orthogonal one is strongly suppressed or removed (e.g., in 2D materials), then we can approximate Eq.~\eqref{eq:alphasol} for the dominant component as $\alpha^{\nu\nu}_j=p_0 b_{j,\nu}/(E_{0,\nu}^{\rm ext}+p_0\sum_i \mathcal{G}^{\nu\nu}_{ji} b_{i,\nu})$ and set $p_0=E_{0,\nu}^{\rm ext}(b_{\rm max}/\alpha_{\rm max}-\sum_i \mathcal{G}^{\nu\nu}_{ji} b_{i,\nu})^{-1}$, while taking the polarizability (and the dipole moment) along the other direction to be zero. Under such conditions, the equation above leads to an accurate mapping of $\alpha_j^{\nu\nu}$ without introducing any inconsistency.
\end{itemize}

\subsection{Dipolar response of electrostatic structures}

\paragraph*{Thin nanodisks.} The in-plane polarizability of a disk with diameter $D\ll\lambda$ and thickness $t\ll D$ composed of a material described by a 2D surface conductivity $\sigma(\omega)$ at frequency $\omega=2\pi c/\lambda$ and embedded in vacuum can be written in the electrostatic limit as \cite{paper257,paper303}
\begin{align}
  \alpha(\omega) = D^3 \sum_m \frac{\zeta_m^2}{1/\eta(\omega)-1/\eta_m},
  \label{eq:polPWF}
\end{align}
where $\eta(\omega)=\ii\sigma(\omega)/D\omega$, and the $m$-sum spans the different multipolar modes supported by the system, parameterized by the eigenvalues $\eta_m$ and mode dipole moments ${\zeta}_m$. For simplicity, in the examples presented in this paper, we take only the dipolar response ($m=1$) and use the fits 
$\eta_1=a_\eta \exp(b_\eta x)+c_\eta$ and $\zeta_1=a_\zeta \exp(b_\zeta x)+c_\zeta$ 
reported in Ref.~\citenum{paper303}, where $x=t/D$. Specifically, for a disk of diameter $D$, we take $a_\eta=0.03801$, $b_\eta=-8.569$, $c_\eta=-0.1108$, $a_\zeta=-0.01267$, $b_\zeta=-45.34$, and $c_\zeta=0.8635$.

The absorption of incident radiation by such a disk is described by its corresponding absorption cross section $\sigma_{\rm abs}(\omega)=\sigma_{\rm ext}(\omega)-\sigma_{\rm sc}(\omega)$, written in terms of the extinction $\sigma_{\rm ext}(\omega)=4\pi k\, {\rm Im}\{\alpha(\omega)\}$ and scattering $\sigma_{\rm sc}(\omega)=8\pi k^4|\alpha(\omega)|^2/3$ counterparts (with $k=\omega/c$) valid when $D\ll\lambda$.

\paragraph*{Thin nanoribbons.} Unlike disks and other finite-size structures, nanoribbons are infinite along their longitudinal direction, and therefore, the associated normal modes are characterized by a wave vector $q$. In the Appendix, we show that, when exposed to an external field component of transversal wave vector $q$, the ribbon exhibits a dipole moment density $\Pb_q$ associated with an effective polarizability tensor $\alpha_q$ given in Eq.~(S8) in the Appendix. For $q=0$, as considered in our results, the response of a dipole can be simply described by its $xx$ component with the form
\begin{align}
  \alpha_0^{xx}(\omega) = W^2 \sum_m \frac{\zeta_{m0}^2}{1/\eta(\omega)-1/\eta_{m0}},
  \label{eq:polribbon}
\end{align}
where $\zeta_{1,0}=0.942$ and $\eta_{1,0}=-0.069$ for the nanoribbon dipolar $m=1$ mode (see Appendix).

\paragraph*{Nanorods.} We model the polarizability of nanorods in vacuum with a permittivity $\epsilon$, length $L$, and tip radius of $L/2R$ (where $R$ is the aspect ratio of the rod),  in the electrostatic limit ($L\ll\lambda$), as \cite{paper300}
\begin{align}
 \alpha(\omega) = \frac{1}{4\pi} \sum_m V_m\left(\frac{1}{\epsilon-1}-\frac{1}{\epsilon_m-1}\right)^{-1},
\end{align}
where, as above, the $m$-sum spans the multipolar modes of the system. For simplicity, we take only the dipolar mode ($m=1$), for which we use fitted parameters from Ref.~\citenum{paper300}, specifically, $V_1=0.896V$ and $\epsilon_1=-1.73R^{1.45}-0.296$, with $V=\pi L^3 (3R-1)/12R^3$ being the nanorod volume.

\subsection{Material modeling}

Here, we show the adopted models to describe the optical response of the different materials considered in this work. For the thin VO$_2$ disks and Au ribbons of small thickness $t$, we assume the structures to be effectively two-dimensional, and we describe them through a surface conductivity related to their permittivity $\epsilon$ by $\sigma=\ii\omega t(1-\epsilon)/4\pi$ \cite{AM1976}. 

\paragraph*{Vanadium dioxide.} For a given value of the metallic fraction $f_{\rm m}$ ranging between $0$ and $1$, we model the mixed-phase permittivity of VO$_2$, $\epsilon_{\rm VO_2}(\omega)$, at frequency $\omega$, using the Bruggeman effective-medium relation \cite{B1935,JFT06,QBC07}
\begin{align}
f_{\rm m}\frac{\epsilon_{\rm m}-\epsilon_{\rm VO_2}}{\epsilon_{\rm m}+2\epsilon_{\rm VO_2}}
+(1-f_{\rm m})\frac{\epsilon_{\rm i}-\epsilon_{\rm VO_2}}{\epsilon_{\rm i}+2\epsilon_{\rm VO_2}}=0,
\label{eq:bruggeman}
\end{align}
where $\epsilon_{\rm i}(\omega)$ and $\epsilon_{\rm m}(\omega)$ are the insulating and metallic phase permittivities of VO$_2$ \cite{BBL20}, respectively [see Fig.~\ref{figS3}(b) in the Appendix]. 

As described in detail in the Appendix, we model the fluence dependence of $f_{\rm m}$ for a VO$_2$ disk with diameter $D=250$~nm and thickness $t=2$~nm (see Fig.~\ref{fig3}) under illumination by a light pulse of fluence $F$ and wavelength $632$~nm as
\begin{align}
 f_{\rm m}(F) = \frac{1}{1+\exp[-(F-F_{\rm m})/\Delta F]},
 \label{eq:fcF}
\end{align}
where $F_{\rm m}=0.395~\mathrm{J/cm^2}$ and $\Delta F=0.019~\mathrm{J/cm^2}$ (see Appendix for details). We note that the relatively large fluence necessary to produce mild temperature increases in the disk reflects its very small absorption cross section [see Fig.~\ref{figS3}(d) in the Appendix].

\paragraph*{Graphene.} We take graphene to be described by an isotropic 2D optical conductivity with the Drude form \cite{paper235}
\begin{align}
  \sigma(\omega)=\frac{\ii e^2}{\pi\hbar^2}\frac{\EF}{\omega+\ii\tau^{-1}},
  \label{eq:graphenesigma}
\end{align}
at frequency $\omega$, where $\EF$ is its Fermi level, 
$\tau=\mu\EF/e\vF^2$ represents the inelastic scattering time for a certain mobility $\mu$, and $\vF\approx c/300$ is the Fermi velocity of electrons in graphene. In this work, we take $\mu=10,000~\mathrm{cm^2/Vs}$. The expression above contains exclusively the intraband response of graphene and, therefore, is valid for $\hbar\omega \lesssim 2\EF$ and $\EF\gg\kB T$ at a certain electron temperature $T$. While here we choose to use Eq.~\eqref{eq:graphenesigma} for the sake of simplicity, a full description of the graphene conductivity including interband transitions, high electron temperature, and/or nonlocal effects can be found in Ref.~\citenum{GP16}.

\paragraph*{Noble metals.} We model the permittivity of gold and silver, used in Figs.~\ref{figS4} and \ref{figS5} in the Appendix, through the Drude model \cite{AM1976},
\begin{align}
  \epsilon(\omega) = \epsilon_{\rm b} - \frac{\omega_{\rm p}^2}{\omega(\omega+\ii\gamma_{\rm m})},
\end{align}
where the plasma frequency $\omega_{\rm p}$, inelastic damping rate $\gamma_{\rm m}$, and background permittivity $\epsilon_{\rm b}$ are fitted from experimental data \cite{JC1972,paper300}. Specifically, for gold, we take $\hbar\omega_{\rm p}=9.06$~eV, $\hbar\gamma_{\rm m}=71$~meV, and $\epsilon_{\rm b}=9.5$; for silver, we use $\hbar\omega_{\rm p}=9.17$~eV, $\hbar\gamma_{\rm m}=21$~meV, and $\epsilon_{\rm b}=4.0$.

\appendix

\section{Electron interacting with an array of ribbons}

\subsection{Induced dipole on a single ribbon}

We follow the so-called plasmon wave function formalism \cite{paper228,paper257,paper303}, which applies to the specific case of a two-dimensional (2D) ribbon excited by an electron beam \cite{paper364,paper430}. We assume that the ribbon has a finite width $W$ along the $x$ direction, extends infinitely along the $y$-direction at the $z=0$ plane, and is described by a local optical conductivity $\sigma(\omega)$ at frequency $\omega$. Within the quasistatic approximation, the electric field $\Eb(\Rb,\omega)$ at the 2D ribbon coordinate $\Rb=(x,y)$ under illumination by a field $\Eb^{\rm ext}(\Rb,\omega)$ can be written self-consistently as
\begin{align}
  \vcalE(\vtheta) = \vcalE^{\rm ext}(\vtheta) + \eta(\omega) \int {\rm d}^2\vtheta\, \mathbf{M}(\vtheta,\vtheta')\cdot\vcalE(\vtheta') ,
  \label{eq:PWF_SC1}
\end{align}
where we define $\vtheta=\Rb/W=(\theta_x,\theta_y)$, $\vcalE(\vtheta)=W\Eb(\vtheta,\omega)$, $\vcalE^{\rm ext}(\vtheta)=W\Eb^{\rm ext}(\vtheta,\omega)$, and $\mathbf{M}(\vtheta,\vtheta')=(\nabla_{\vtheta} \otimes \nabla_{\vtheta})\, |\vtheta-\vtheta'|^{-1}$. Due to the translational invariance of the system along $y$, it is convenient to expand the fields as
\begin{align}
 \vcalE(\vtheta)=\frac{1}{2\pi} \int {\rm d}\q\, \vcalE_{\q}(\theta_x) \ee^{\ii \q \theta_y} ,
\end{align}
where each $\q=q W$ denotes a normal mode of the system with wave vector $q$ along $\yy$. By doing so, we can rewrite Eq.~\eqref{eq:PWF_SC1} as
\begin{align}
  \vcalE_\q(\theta_x) = \vcalE_\q^{\rm ext}(\theta_x) + \eta(\omega) \int_{-1/2}^{1/2} {\rm d}\theta_x\, \tilde{\mathbf{M}}_\q(\theta_x,\theta_x')\cdot\vcalE_\q(\theta_x') ,
  \label{eq:PWF_SC2}
\end{align}
where the operator $\tilde{\mathbf{M}}_\q(\theta_x,\theta_x')=2(\nabla_{\vtheta} \otimes \nabla_{\vtheta})K_0(|\q||\theta_x-\theta_x'|)$ is expressed in terms of the modified Bessel function $K_0$~\cite{paper364}, $\vcalE_\q^{\rm ext}(\theta_x) = \int {\rm d}\theta_y\, \vcalE^{\rm ext}(\vtheta) \ee^{-\ii \q\theta_y}$.

In the absence of an external field, Eq.~\eqref{eq:PWF_SC2} becomes an eigenvalue problem,
\begin{align}
\vcalE_{\j\q}(\theta_x) = \eta_{\j\q} \int_{-1/2}^{1/2} {\rm d}\theta_x'\, \tilde{\mathbf{M}}_\q(\theta_x,\theta_x')\cdot\vcalE_{\j\q}(\theta_x'),
\end{align}
with eigenvalues $\eta_{\j\q}$ and associated eigenvectors $\vcalE_{\j\q}(\theta_x)$ here presumed to be known (see discussion below). Using this relation, we can readily find that the $\q$-component of the self-consistent total electric field in Eq.~\eqref{eq:PWF_SC2} within the ribbon can be written as
\begin{align}
  \vcalE_{\q}(\theta_x) = \sum_\j \frac{c_{\j\q}}{1-\eta(\omega)/\eta_{\j\q}} ,
  \label{eq:Ejq}
\end{align}
with coefficients
\begin{subequations}
\begin{align}
  c_{\j\q} &= \int_{-1/2}^{1/2} {\rm d}\theta_x\, \vcalE_\q^{\rm ext}(\theta_x) \cdot \vcalE_{\j\q}(\theta_x) .
  \label{eq:cjq} 
\end{align}
We note that, in general, $\vcalE^{\rm ext}(\vtheta)=-\nabla_{\vtheta}\,\Phi^{\rm ext}(\vtheta)$ or, equivalently, $\vcalE_\q^{\rm ext}(\vtheta)=-\nabla_{\theta_x}\Phi_\q^{\rm ext}(\theta_x)$, where we introduce $\nabla_{\theta_x}=(\partial_{\theta_x},\ii \q)$, with $\Phi_\q^{\rm ext}(\theta_x)=\int{\rm d}\theta_y \Phi^{\rm ext}(\vtheta)\ee^{-\ii \q \theta_y}$ being the electrostatic potential associated with the external radiation. This allows us to evaluate the integral in Eq.~\eqref{eq:cjq} by parts to express it in the alternative form
\begin{align}
  c_{\j\q} = \int_{-1/2}^{1/2} {\rm d}\theta_x\, \Phi_\q^{\rm ext}(\theta_x) \rho_{\j\q}(\theta_x) ,
  \label{eq:cjq2}
\end{align}
\end{subequations}
given in terms of the plasmon wave functions $\rho_{\j\q}(\theta_x)=\nabla_{\theta_x}\cdot\vcalE_{\j\q}(\theta_x)$ . 

Using the continuity equation $\ii\omega\rho^{\rm ind}(\Rb)=\nabla\cdot \Jb(\Rb)$ and Ohm's law $\Jb(\Rb)=\sigma(\omega)\Eb(\Rb)$, the induced charge on the ribbon can be found as $\rho^{\rm ind}(\vtheta)=(2\pi)^{-1} \int{\rm d}\q\, \rho_\q^{\rm ind}(\theta_x) \ee^{\ii \q \theta_y}$, where 
\begin{align}
 \rho_\q^{\rm ind}(\theta_x) = \frac{1}{W} \sum_\j \frac{c_{\j\q}}{1/\eta_{\j\q}-1/\eta(\omega)} \rho_{\j\q}(\theta_x) .
 \label{eq:rhoind}
\end{align}
From here, the induced dipole moment density along the ribbon length can be written as $\pb(\theta_y) = \int_{-1/2}^{1/2} {\rm d}\theta_x\, \rho^{\rm ind}(\vtheta)\vtheta$. By using Eq.~\eqref{eq:rhoind} and employing a Fourier transform to bring the induced dipole moment density into the $\q$-space, $\pb_\q=\int{\rm d}\theta_y\, \pb(\theta_y)\ee^{-\ii \q \theta_y}$, we obtain the components
\begin{subequations}\label{eq:pind}
\begin{align}
p_{\q,x} &= W^2 \int_{-1/2}^{1/2} {\rm d}\theta_x\, \rho^{\rm ind}_\q(\theta_x)\theta_x \nonumber\\ &= W  \sum_\j \frac{c_{\j\q} }{1/\eta_{\j\q}-1/\eta(\omega)} \zeta_{\j\q} , \\
p_{\q,y} &= \ii W^2 \partial_\q \int_{-1/2}^{1/2} {\rm d}\theta_x\, \rho^{\rm ind}_\q(\theta_x) \nonumber\\ &= \ii W  \partial_\q \sum_\j \frac{c_{\j\q}}{1/\eta_{\j\q}-1/\eta(\omega)} \chi_{\j\q} ,
\end{align}
\end{subequations}
where we define $\zeta_{\j\q}=\int_{-1/2}^{1/2} {\rm d}\theta_x\,  \rho_{\j\q}(\theta_x) \theta_x$ and $\chi_{\j\q}=\int_{-1/2}^{1/2} {\rm d}\theta_x\, \rho_{\j\q}(\theta_x)$, with the operator $\partial_\q$ arising from an integration by parts. Importantly, $\chi_{\j\q}$ vanishes when $\q=0$ due to charge neutrality, but the same is not true when $\q \neq 0$. 

The combination of Eqs.~\eqref{eq:pind} with Eq.~\eqref{eq:cjq} determines the dipole density along $y$ induced on the ribbon by an incident electric field $\q$-component $\vcalE_\q^{\rm ext}(\theta_x)$. However, when the width $W$ is much smaller than the wavelength $\lambda=2\pi c/\omega$, we can neglect the variation of the incident field across the ribbon width and approximate $\vcalE_\q^{\rm ext}(\theta_x) \approx \vcalE_\q^{\rm ext}(0) \equiv \vcalE_\q^{0}=(\calE_{\q,x}^0,\calE_{\q,y}^0)$. Under such conditions, we can write $\Phi^{\rm ext}_\q(\theta_x)\approx-[\theta_x \calE_{\q,x}^0 + (\ii \q)^{-1} \calE_{\q,y}^0]$, from where we obtain from Eq.~\eqref{eq:cjq2} that $c_{\j\q}=-\zeta_{\j\q} \calE_{\q,x}^0 + (\ii/\q) \chi_{\j\q} \calE_{\q,y}^0$. Plugging this result into Eqs.~\eqref{eq:pind} and noting that we can write, in general, $\pb_\q = \dvec\alpha_\q \cdot\vcalE_\q^{0}/W$, we can finally define the $\q$-component polarizability tensor $\dvec\alpha_\q$ as
\begin{align}
\dvec\alpha_\q = W^2 \sum_\j
\begin{bmatrix}
 \zeta_{\j\q}^2 & \frac{-\ii}{\q} \zeta_{\j\q}\chi_{\j\q} \\
 \ii \partial_\q \zeta_{\j\q}\chi_{\j\q} & \partial_\q \frac{1}{\q}\chi_{\j\q}^2
\end{bmatrix} 
 \frac{1}{\eta(\omega)^{-1}-\eta_{\j\q}^{-1}} .
\end{align}
We note that the $\partial_\q$ operators act on every $\q$-function to the right-hand side of them. For $\q=0$, only the tensor component $xx$ is nonzero and takes the form of Eq.~\eqref{eq:polribbon}.

Using a Chebyshev polynomials basis described in Ref.~\citenum{paper364}, the eigenmodes of the system can be expanded as
\begin{align}
 \vcalE_{\j\q}(\theta_x) = \sum_{n=0}^{\infty} \sqrt{1-4\theta_x^2}U_n(2\theta_x)(u_{\j\q,n},-\ii v_{\j\q,n}) ,
\end{align}
where $U_n$ denotes the Chebyshev polynomial of the second kind and $u_{\j\q,n}$ and $v_{\j\q,n}$ are expansion coefficients that can be determined as described in the mentioned reference, together with its associated eigenvalues $\eta_{\j\q}$. In this basis, the plasmon wave functions are written as
\begin{align}
  \rho_{\j\q}(\theta_x)=\sum_{n=0}^{\infty} \Bigg[&-2u_{\j\q,n}(n+1)\frac{T_{n+1}(2\theta_x)}{\sqrt{1-4\theta_x^2}} \nonumber \\ & +\q v_{\j\q,n} \sqrt{1-4\theta_x^2} U_n(2\theta_x) \Bigg] ,
\end{align}
where $T_n$ is a Chebyshev polynomial of the first kind. From the orthogonality conditions of these polynomials,
\begin{subequations}
\begin{align}
 \int_{-1}^{1} dx\, \frac{T_n(x)T_m(x)}{\sqrt{1-x^2}}&=\frac{\pi}{2} \delta_{nm}(1+\delta_{n0}) , \\
 \int_{-1}^{1} dx\, U_n(x)U_m(x)\sqrt{1-x^2}&=\frac{\pi}{2}\delta_{nm} ,
\end{align}
\end{subequations}
and noting that $T_0(x)=U_0(x)=1$ and $T_1(x)=U_1(x)/2=x$, we can evaluate the integrals of $\rho_{\j\q}(\theta_x)$ across the ribbon to derive
\begin{subequations}
\begin{align}
  \zeta_{\j\q} &= -\frac{\pi}{4} u_{\j\q,0}+\q \frac{\pi}{16} v_{\j\q,1} , \\
  \chi_{\j\q} &= \q \frac{\pi}{4} v_{\j\q,0}.
\end{align}
\end{subequations}
As anticipated, $\chi_{\j\q}$ vanishes for $\q=0$, and we obtain $\zeta_{\j0}=-(\pi/4)u_{\j\q,0}$. 

The coefficients $u_{\j\q,n}$ and $v_{\j\q,n}$, along with their associated eigenvalues $\eta_{\j\q}$, can be found by the method described in Ref.~\citenum{paper364}, and for $\q=0$ are tabulated in Tables S1 and S2 of its Supplementary Information. For the dipolar mode in particular ($\j=1$), we have $\eta_{1,0}=-0.069$ and $u_{1,0,0}=-1.200$, yielding $\zeta_{1,0}=0.942$. These parameters coincide with those reported in Ref.~\citenum{paper257} for a ribbon illuminated by a plane wave, since the $\q=0$ component of an electron beam field decomposition is equivalent to a normally-impinging plane wave.

\subsection{Line dipole--line dipole interaction}

We consider a line extending along $\yy$ and intersecting the point $\rb_{\perp}'=(x',0,z')$, carrying a dipole moment density $\Pb(y)$ per unit length along the $y$-direction. The field generated by such line at position $\rb=(x,y,z)$ can be written as
\begin{align}
 \Eb^{\rm dip}(\rb)=\int{\rm d}y'\, (k^2+\nabla_{\rb}\otimes\nabla_{\rb})\frac{\ee^{\ii k |\rb-\rb'|}}{|\rb-\rb'|} \cdot \Pb(y') ,
\end{align}
with $\rb'=\rb_{\perp}'+y' \yy$, $\nabla_{\rb}=(\partial_x,\partial_y,\partial_z)$, and we have used the point dipole--point dipole interaction Green's tensor [see Eq.~\eqref{eq:pj}]. Now, similarly to the previous section, we employ a Fourier transform to write $\Pb(y)=(2\pi)^{-1} \int{\rm d}q\, \Pb_q \ee^{\ii q y}$. Replacing this in the equation above and using the identity
\begin{align}
 \int{\rm d}y' \frac{\ee^{\ii k |\rb-\rb'|}}{|\rb-\rb'|}\ee^{\ii q y'} = \frac{\ii}{4}H_0^{(1)}(\kappa |\rb_{\perp}-\rb_{\perp}'|)\ee^{\ii q y} ,
\end{align}
expressed in terms of the Hankel function $H_0^{(1)}$, with $\kappa=\sqrt{k^2-q^2}$ and $\rb_{\perp}=(x,0,z)$, we readily find that $\Eb^{\rm ind}_q(\rb_{\perp})=\int{\rm d}y\, \Eb^{\rm ind}(\rb)\ee^{-\ii q y}=\mathcal{G}_q(\rb_{\perp},\rb_{\perp}')\cdot \Pb_q$, such that the line dipole--line dipole interaction Green's tensor associated with wave vector component $q$ can be written as
\begin{align}
  \mathcal{G}_q(\rb_{\perp},\rb_{\perp}')=\frac{\ii}{4}(k^2+\nabla_{\perp}\otimes\nabla_{\perp})H_0^{(1)}(\kappa |\rb_{\perp}-\rb_{\perp}'|) ,
\end{align}
where $\nabla_{\perp}=(\partial_x,\ii q,\partial_z)$. By explicitly performing the derivations, the Green tensor can be expressed in the closed form
\begin{align}
\mathcal{G}_q(\rb_{\perp},\rb_{\perp}')=\frac{\ii \kappa^2}{4}\Bigg[&C_{\rho}(\hat{\mathbf{\rho}}\otimes\hat{\mathbf{\rho}})+C_{\phi}(\hat{\mathbf{\phi}}\otimes\hat{\mathbf{\phi}}) + \frac{k^2}{\kappa^2} H_0(\yy \otimes\yy)  \nonumber \\ &-\ii \frac{q}{\kappa} H_1(\hat{\mathbf{\rho}}\otimes\yy+\yy\otimes\hat{\mathbf{\rho}}) \Bigg] ,
\end{align}
where $\vec\rho=\rb_{\perp}-\rb_{\perp}'$, $\hat{\mathbf{\rho}}=\vec\rho/\rho$, $\rho=|\vec\rho|$, $\hat{\mathbf{\phi}}=\yy\times\hat{\mathbf{\rho}}$, $H_\nu=H_\nu^{(1)}(\kappa \rho)$, and we introduce
\begin{align}
  C_{\rho}&=\left(\frac{k^2}{\kappa^2}+\frac{3}{2}\right)H_0-\frac{1}{2}H_2 , \\
  C_\phi&=\left(\frac{k^2}{\kappa^2}-\frac{1}{2} \right)H_0+\frac{1}{2}H_2 .
\end{align}
For the results presented in this paper, we are primarily interested in the $xx$ component of the tensor for $q=0$ and for $z=z'=0$, for which we find
\begin{align}
 \mathcal{G}_0^{xx}(x,x')=\frac{\ii k^2}{2}\Big[&H_0^{(1)}(k|x-x'|)\nonumber \\ &-\frac{1}{2k|x-x'|}H_1^{(1)}(k|x-x'|)\Big] .
\end{align}

\section{Vanadium dioxide phase-changing dynamics}

The insulator--metal transition (IMT) in VO$_2$ is triggered by an increase in temperature $T$, with the metallic fraction $f_{\rm m}$ being well described by a sigmoidal law 
\begin{align}
f_{\rm m}(T) = \frac{1}{1+\exp[-(T-T_{\rm m})/\Delta T]},
\label{eq:fT}
\end{align}
where $T_{\rm m}\approx 341$~K and $\Delta T\approx 2$~K \cite{JFT06} (see Fig.~\ref{figS3}(c)). This dependence reflects the dynamics of the phase-change: little variation below the IMT window, followed by a rapid rise once nucleation, growth, and percolation of metallic domains activate, and saturation as the remaining insulating volume is exhausted \cite{JFT06,QBC07,VBB68}. As the temperature decreases, the material relaxes back toward the insulating state. Because the IMT in VO$_2$ is a first-order phase transition, it exhibits thermal hysteresis \cite{VBB68,GRO18,HHG19}, so the parameters stated above for Eq.~\eqref{eq:fT} do not describe the cooling branch (which is also sigmoidal, but with its own onset, midpoint, and width). However, keeping hystersis into account, complete tuning can be achieved with an appropriate heating/cooling protocol. 

To estimate the temperature reached by a small VO$_2$ sample with volume $V$ and absorption cross section $\sigma_{\rm abs}(\omega_{\rm p})$ under pumping by a short light pulse with frequency $\omega_{\rm p}$ and fluence $F_0$, we solve the adiabatic energy-balance \cite{VC81}
\begin{align}
F_0=V \int_{T_0}^T dT'\frac{c_{\rm V}}{\sigma_{\rm abs}(\omega_{\rm p})},
\end{align}
with $T_0$ being the ambient temperature, and $c_{\rm V}\approx3.0$~J/cm$^{3}$K \cite{OKR10} the heat capacity per unit volume of VO$_2$ (which we take as approximately independent of $T$ for the purpose of this estimation). We note that $\sigma_{\rm abs}$ depends on temperature via the variation in permittivity of the sample at the pump wavelength as the IMT occurs. Applying this expression to VO$_2$ disk with the characteristics specified in Fig.~\ref{fig3}, we obtain a fluence-dependent $f_{\rm m}$ that can be accurately described by with a sigmoidal function with the form shown in Fig.~\ref{figS3}(d), described by Eq.~\eqref{eq:fcF} (with the parameters therein indicated).

For higher accuracy, the simple temperature estimate above can be replaced by a time-resolved thermal model (eg. the two-temperature model \cite{A1939,FR06,PKE11}), including an accurate temperature-dependence of the heat capacity $c_{\rm V}(T)$. The resulting $T(t)$ curve can then be used with first-order IMT kinetics, namely the Johnson-Mehl-Avrami-Kolmogorov (JMAK) growth model \cite{A1939,FR06} in combination with an Arrhenius temperature dependence \cite{FR06,CHK22}, to determine the final metallic fraction for a given pump pulse characteristics.

\section{Supplementary figures}

\begin{figure*}[htbp]
    \centering
    \includegraphics[width=\linewidth]{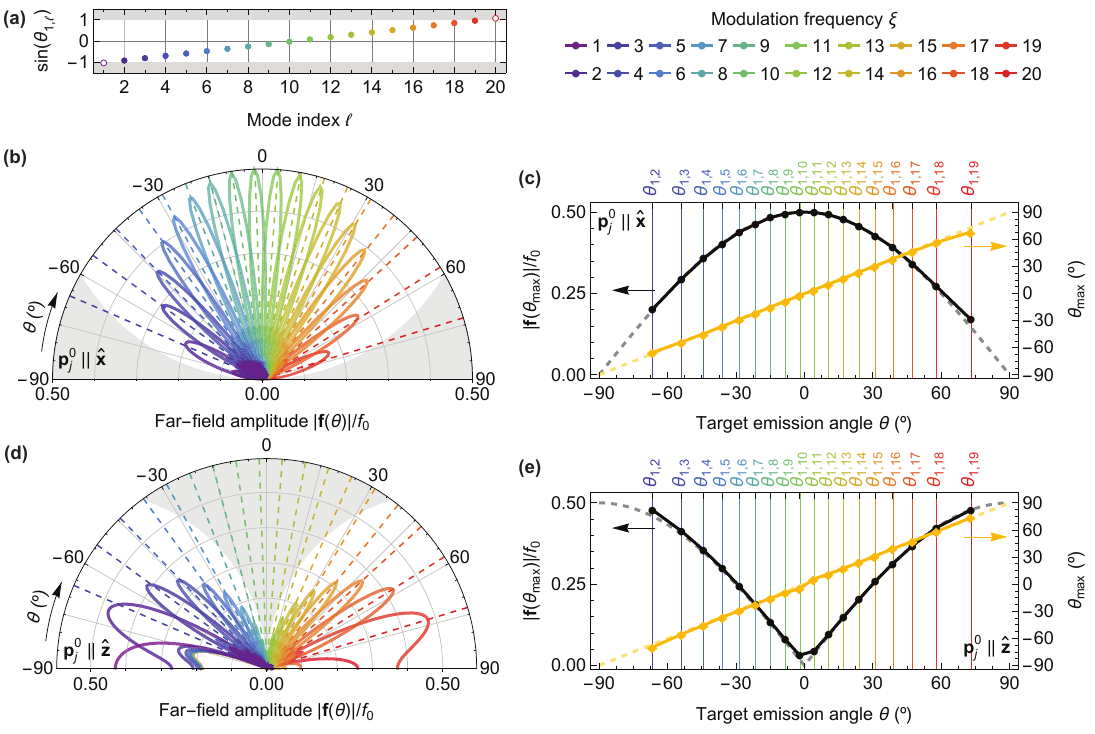}
    \caption{\textbf{Cathodoluminescence steering with large non-uniform arrays.} \textbf{(a)} GSP condition in Eq.~\eqref{eq:SPcondition} for $n=1$ and $\ell=1-20$ in an array with $N=101$ elements and period $a/\lambda = 0.090$, excited by an electron beam with velocity $v=0.1c$ ($\approx 2.6$~keV). Modes $\ell=2-19$ lie within the region $|\sin \theta_{1,\ell}|\leq 1$ and are marked by filled circles, whereas modes $\ell=1$ and $\ell=20$ lie outside the same region and are marked with empty circles. \textbf{(b)} Angle-resolved cathodoluminescence far-field emission amplitude $|\fb(\theta)|$ polar plot, normalized to $f_0=Nk^2p_0$, as a function of $\theta$ angle in the $\phi=0$ plane, for an electron passing over an array whose induced dipole moments are given by Eq.~\eqref{eq:pjharmonic} with $\xi$ as indicated in the legend and polarized along $x$ ($\pb_j^0 \parallel \xx$). The colored dashed lines mark the position of the target angles $\theta_{1,\xi}$ for $\xi=2-19$ and the shaded gray area represents the condition $|\fb(\theta)|/f_0 > \cos(\theta)$. \textbf{(c)} Peak emission angle $\theta_{\rm max}$ (right axis) and corresponding peak far-field amplitude $|\fb(\theta_{\rm max})|/f_0$ (left axis) as a function of the target emission frequency $\theta$ for $\xi=2-19$ (see top axis). The dashed gray curve represents the function $|\fb(\theta)|/f_0=\cos(\theta)/2$ and the yellow dashed line represents the condition $\theta_{\rm max}=\theta$. \textbf{(d,e)} Same as (b,c), respectively, but for dipoles polarized along $z$ ($\pb_j^0 \parallel \zz$). In (d), the shaded gray area represents the condition $|\fb(\theta)|/f_0 > \sin(\theta)$. In (e), the dashed gray curve represents the function $|\fb(\theta)|/f_0=\sin(\theta)/2$.}
    \label{figS1}
\end{figure*}

\begin{figure*}[htbp]
    \centering
    \includegraphics[width=0.923\linewidth]{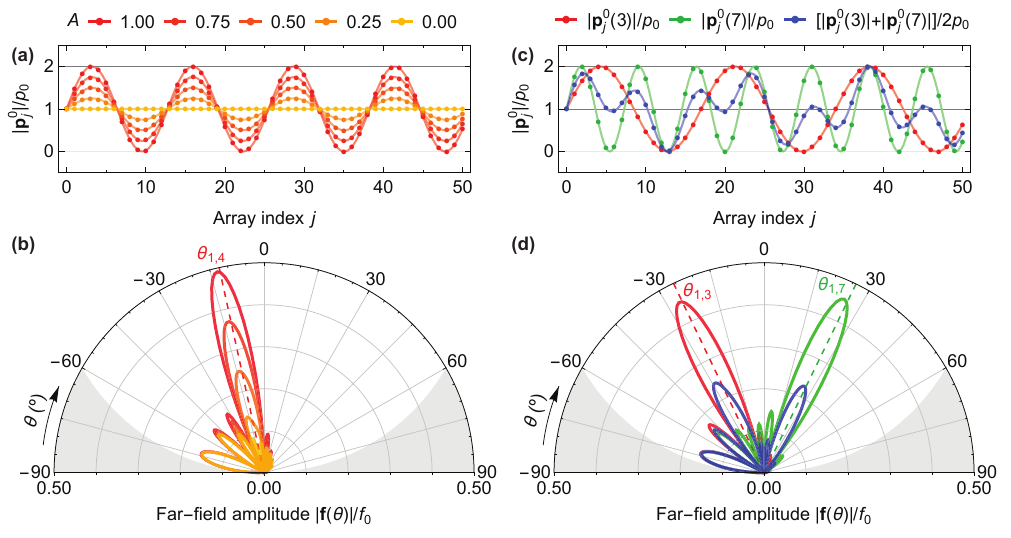}
    \caption{\textbf{Properties of the CL far-field emission profile.} \textbf{(a)} Induced dipole distribution profile following Eq.~\eqref{eq:pjharmonic} with $\xi=4$, with the parameter $A$ ranging from 0 to 1 (see legend). \textbf{(b)} CL angle-resolved far-field distribution $|\fb(\theta)|$ at $\phi=0$ for each of the color-coded distributions in (a). \textbf{(c,d)} Same as (a,b), respectively, but for induced dipole profiles following Eq.~\eqref{eq:pjharmonic} with $A=1$ for $\xi=3$, $\xi=7$, and the averaged superposition of both profiles, as shown in the legend.}
    \label{figS2}
\end{figure*}

\begin{figure*}[htbp]
    \centering
    \includegraphics[width=\linewidth]{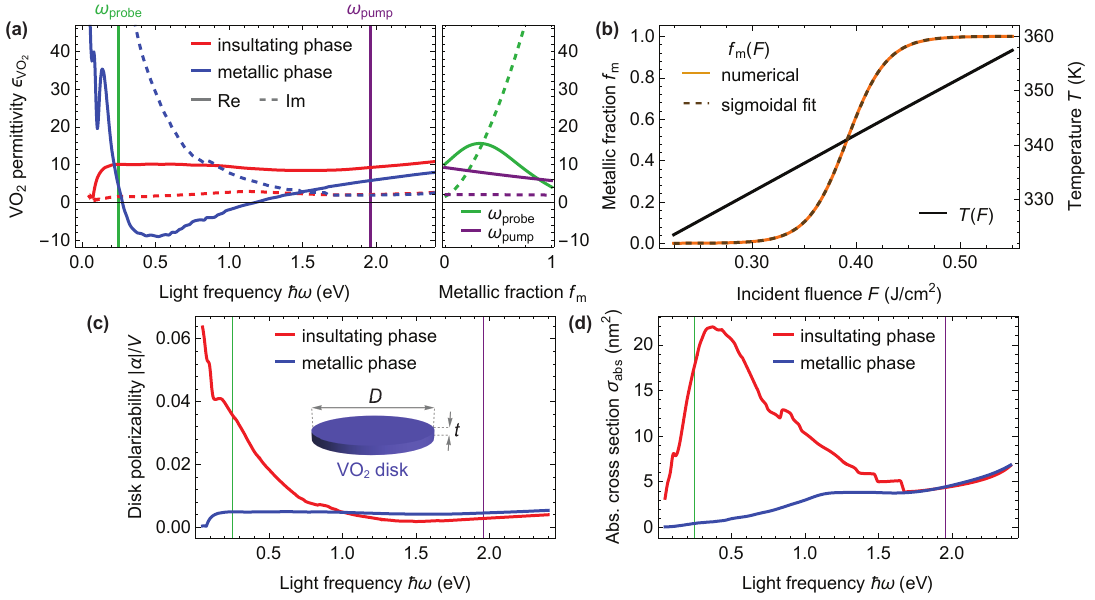}
    \caption{\textbf{VO$_2$ thermo-optical properties.} \textbf{(a)} Real and imaginary parts of the permittivity of VO$_2$ in its insulating ($f_{\rm m}=0$) and metallic ($f_{\rm m}=1$) phases. On the right panel, we show the variation of the permittivity with $f_{\rm m}$ for two representative frequencies $\hbar\omega_{\rm probe}=0.25$~eV and $\hbar\omega_{\rm pump}=1.96$~eV, marked by vertical lines on the left plot. \textbf{(b)} Variation of the temperature (right axis) and corresponding metallic fraction $f_{\rm m}$ (left axis) of a VO$_2$ disk with $D=250$~nm and $t=2$~nm (see inset in (c)) as a function of incident fluence $F$ at $\omega_{\rm pump}$. \textbf{(d)} Disk polarizability $|\alpha|$ (normalized to its volume $V=\pi D^2 t/4$) and \textbf{(e)} absorption cross section $\sigma_{\rm abs}$, as a function of light frequency, for the insulating and metallic phases.}
    \label{figS3}
\end{figure*}

\begin{figure*}[htbp]
    \centering
     \includegraphics[width=0.85\linewidth]{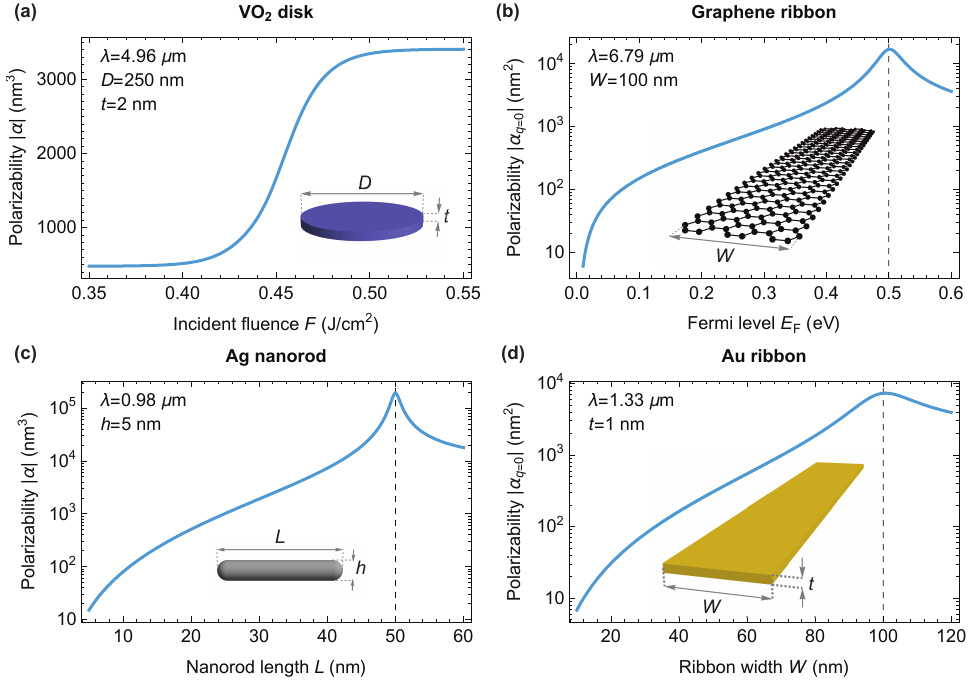}
    \caption{\textbf{Polarizability tuning.} \textbf{(a)} Variation of the polarizability amplitude $|\alpha|$ of a VO$_2$ disk with diameter $D$ and thickness $t$ (see scheme and labels), as a function of incident fluence $F$ for pumping at $632$~nm. \textbf{(b)} Variation of the effective polarizability at $q=0$ (see Methods and Appendix) of a graphene ribbon with width $W$, as a function of Fermi energy $\EF$. \textbf{(c,d)} Same as (a,b), respectively, but for (c) a silver nanorod of length $L$ and tip diameter $h$, as a function of $L$, and (d) a gold ribbon with thickness $t$ and width $W$ as a function of $W$. In all plots, the wavelength at which the polarizability is calculated is indicated by a label.}
    \label{figS4}
\end{figure*}

\begin{figure*}[htbp]
    \centering
    \includegraphics[width=\linewidth]{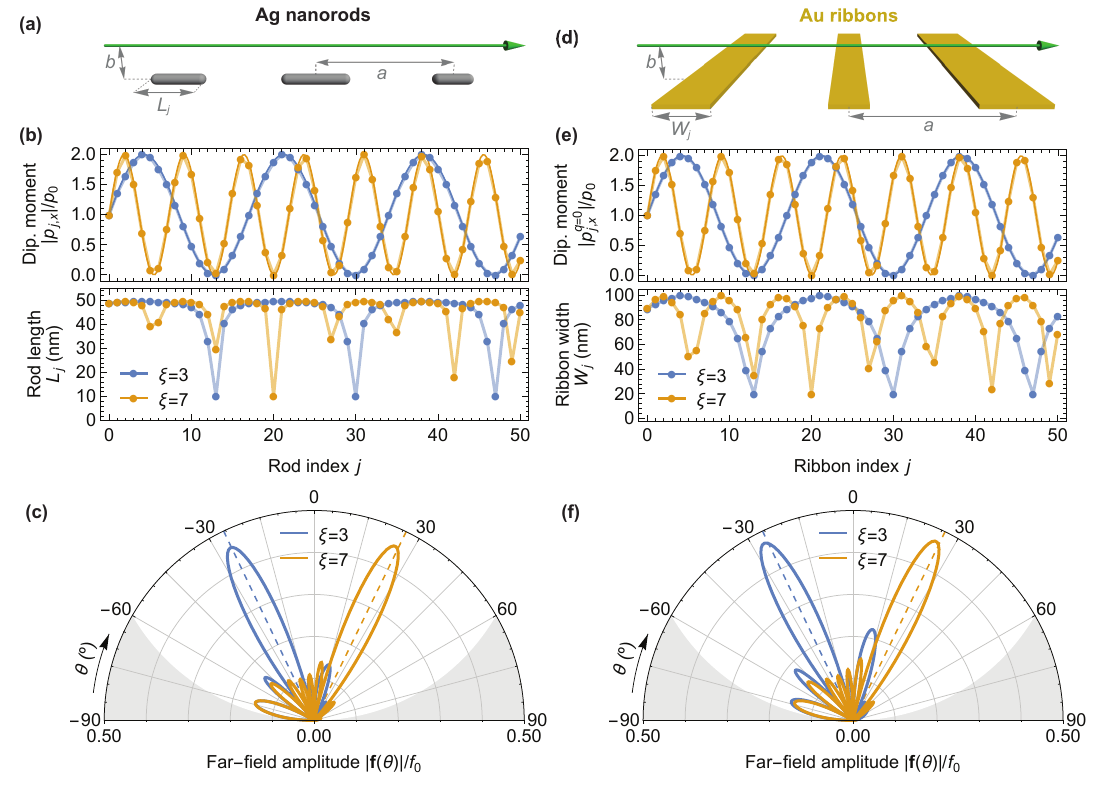}
    \caption{\textbf{Passive tuning of CL emission.} \textbf{(a)} Scheme of an array of silver nanorods with thickness $h=5$~nm and length $L_j$, separated from their nearest neighbors by a center-to-center distance $a=90~$nm, with an electron passing parallel to the array at a distance $b=10$~nm and with velocity $v=0.1c$. \textbf{(b)} Induced dipole moment (top) at wavelength $\lambda=2\pi c/\omega=985$~nm ($\approx 1.26$~eV) on the array, with $p_0=9.45 (eL_{\rm max}/\omega)$, as a function of element $j$, for the color-coordinated rod length distribution $L_j$ (bottom) ranging from $L_{\rm min}=10$~nm to $L_{\rm max}=50$~nm \cite{KSC21}. \textbf{(c)} Far-field emission distribution for the same-color array distributions in (b), with $f_0=50.0 (e/L_{\rm max}\omega)$. \textbf{(d-f)} Same as (a-c), but for an array of gold ribbons with thickness $t=1$~nm and width $W_j$ ranging from $W_{\rm min}=20$~nm to $W_{\rm max}=100$~nm \cite{PTQ24}, with $a=120~\mathrm{nm}$ and $\lambda=1.33~\mathrm{\mu m}$ ($\approx0.93$~eV). In (e), where the dipole moment corresponding to $q=0$ component, with $p_0=6.72(eW_{\rm max}/\omega)$. In (f), we have $f_0=76.1/(e/W_{\rm max}\omega)$.}
    \label{figS5}
\end{figure*}

We include Figs.~\ref{figS1}--\ref{figS5} containing supplementary material to the results presented in the main part of the paper.


\begin{acknowledgments}
E.~J.~C.~D. acknowledges support from the European Union (Marie Skłodowska-Curie Actions, grant agreement No. 101209876). A.~R.~E. and F.~J.~G.~A. acknowledge support from ERC (Advanced Grant 789104-eNANO), the Spanish MINECO (MAT2017-88492-R and SEV2015-0522), the Catalan CERCA Program, and Fundaci\'o Privada Cellex. J.~D.~C. acknowledges support from Independent Research Fund Denmark (grant no. 0165-00051B). The Center for Polariton-driven Light--Matter Interactions (POLIMA) is funded by the Danish National Research Foundation (Project No.~DNRF165).
\end{acknowledgments}

\section*{Contributions}

E.~J.~C.~D. and J.~D.~C. conceived the study. E.~J.~C.~D. performed the calculations, with input and analysis from all authors. All authors contributed to the development of the theory and writing of the manuscript.

\section*{Data availability}

All data needed to evaluate the conclusions in the paper are present in the paper. Additional data related to this paper may be requested from the authors.

\section*{Conflict of interest}

The authors declare no competing interests.


\begin{thebibliography}{61}%
\makeatletter
\providecommand \@ifxundefined [1]{%
 \@ifx{#1\undefined}
}%
\providecommand \@ifnum [1]{%
 \ifnum #1\expandafter \@firstoftwo
 \else \expandafter \@secondoftwo
 \fi
}%
\providecommand \@ifx [1]{%
 \ifx #1\expandafter \@firstoftwo
 \else \expandafter \@secondoftwo
 \fi
}%
\providecommand \natexlab [1]{#1}%
\providecommand \enquote  [1]{``#1''}%
\providecommand \bibnamefont  [1]{#1}%
\providecommand \bibfnamefont [1]{#1}%
\providecommand \citenamefont [1]{#1}%
\providecommand \href@noop [0]{\@secondoftwo}%
\providecommand \href [0]{\begingroup \@sanitize@url \@href}%
\providecommand \@href[1]{\@@startlink{#1}\@@href}%
\providecommand \@@href[1]{\endgroup#1\@@endlink}%
\providecommand \@sanitize@url [0]{\catcode `\\12\catcode `\$12\catcode `\&12\catcode `\#12\catcode `\^12\catcode `\_12\catcode `\%12\relax}%
\providecommand \@@startlink[1]{}%
\providecommand \@@endlink[0]{}%
\providecommand \url  [0]{\begingroup\@sanitize@url \@url }%
\providecommand \@url [1]{\endgroup\@href {#1}{\urlprefix }}%
\providecommand \urlprefix  [0]{URL }%
\providecommand \Eprint [0]{\href }%
\providecommand \doibase [0]{http://dx.doi.org/}%
\providecommand \selectlanguage [0]{\@gobble}%
\providecommand \bibinfo  [0]{\@secondoftwo}%
\providecommand \bibfield  [0]{\@secondoftwo}%
\providecommand \translation [1]{[#1]}%
\providecommand \BibitemOpen [0]{}%
\providecommand \bibitemStop [0]{}%
\providecommand \bibitemNoStop [0]{.\EOS\space}%
\providecommand \EOS [0]{\spacefactor3000\relax}%
\providecommand \BibitemShut  [1]{\csname bibitem#1\endcsname}%
\let\auto@bib@innerbib\@empty
\bibitem [{\citenamefont {Yu}\ and\ \citenamefont {Capasso}(2014)}]{yu2014flat}%
  \BibitemOpen
  \bibfield  {author} {\bibinfo {author} {\bibfnamefont {Nanfang}\ \bibnamefont {Yu}}\ and\ \bibinfo {author} {\bibfnamefont {Federico}\ \bibnamefont {Capasso}},\ }\bibfield  {title} {\enquote {\bibinfo {title} {Flat optics with designer metasurfaces},}\ }\href {\doibase 10.1038/nmat3839} {\bibfield  {journal} {\bibinfo  {journal} {Nat.\ Mater.}\ }\textbf {\bibinfo {volume} {13}},\ \bibinfo {pages} {139--150} (\bibinfo {year} {2014})}\BibitemShut {NoStop}%
\bibitem [{\citenamefont {Hu}\ \emph {et~al.}(2021)\citenamefont {Hu}, \citenamefont {Bandyopadhyay}, \citenamefont {hui Liu},\ and\ \citenamefont {yang Shao}}]{hu2021review}%
  \BibitemOpen
  \bibfield  {author} {\bibinfo {author} {\bibfnamefont {Jie}\ \bibnamefont {Hu}}, \bibinfo {author} {\bibfnamefont {Sankhyabrata}\ \bibnamefont {Bandyopadhyay}}, \bibinfo {author} {\bibfnamefont {Yu}~\bibnamefont {hui Liu}}, \ and\ \bibinfo {author} {\bibfnamefont {Li}~\bibnamefont {yang Shao}},\ }\bibfield  {title} {\enquote {\bibinfo {title} {A review on metasurface: from principle to smart metadevices},}\ }\href {\doibase 10.3389/fphy.2020.586087} {\bibfield  {journal} {\bibinfo  {journal} {Front.\ Phys.}\ }\textbf {\bibinfo {volume} {8}},\ \bibinfo {pages} {586087} (\bibinfo {year} {2021})}\BibitemShut {NoStop}%
\bibitem [{\citenamefont {Li}\ \emph {et~al.}(2017)\citenamefont {Li}, \citenamefont {Zhang},\ and\ \citenamefont {Zentgraf}}]{li2017nonlinear}%
  \BibitemOpen
  \bibfield  {author} {\bibinfo {author} {\bibfnamefont {Guixin}\ \bibnamefont {Li}}, \bibinfo {author} {\bibfnamefont {Shuang}\ \bibnamefont {Zhang}}, \ and\ \bibinfo {author} {\bibfnamefont {Thomas}\ \bibnamefont {Zentgraf}},\ }\bibfield  {title} {\enquote {\bibinfo {title} {Nonlinear photonic metasurfaces},}\ }\href {\doibase 10.1038/natrevmats.2017.10} {\bibfield  {journal} {\bibinfo  {journal} {Nat.\ Rev.\ Mater.}\ }\textbf {\bibinfo {volume} {2}},\ \bibinfo {pages} {17010} (\bibinfo {year} {2017})}\BibitemShut {NoStop}%
\bibitem [{\citenamefont {Keren-Zur}\ \emph {et~al.}(2018)\citenamefont {Keren-Zur}, \citenamefont {Michaeli}, \citenamefont {Suchowski},\ and\ \citenamefont {Ellenbogen}}]{keren2018shaping}%
  \BibitemOpen
  \bibfield  {author} {\bibinfo {author} {\bibfnamefont {Shay}\ \bibnamefont {Keren-Zur}}, \bibinfo {author} {\bibfnamefont {Lior}\ \bibnamefont {Michaeli}}, \bibinfo {author} {\bibfnamefont {Haim}\ \bibnamefont {Suchowski}}, \ and\ \bibinfo {author} {\bibfnamefont {Tal}\ \bibnamefont {Ellenbogen}},\ }\bibfield  {title} {\enquote {\bibinfo {title} {Shaping light with nonlinear metasurfaces},}\ }\href {\doibase 10.1364/AOP.10.000309} {\bibfield  {journal} {\bibinfo  {journal} {Adv.\ Opt.\ Photon.}\ }\textbf {\bibinfo {volume} {10}},\ \bibinfo {pages} {309--353} (\bibinfo {year} {2018})}\BibitemShut {NoStop}%
\bibitem [{\citenamefont {Semmlinger}\ \emph {et~al.}(2018)\citenamefont {Semmlinger}, \citenamefont {Tseng}, \citenamefont {Yang}, \citenamefont {Zhang}, \citenamefont {Zhang}, \citenamefont {Tsai}, \citenamefont {Tsai}, \citenamefont {Nordlander},\ and\ \citenamefont {Halas}}]{semmlinger2018vacuum}%
  \BibitemOpen
  \bibfield  {author} {\bibinfo {author} {\bibfnamefont {Michael}\ \bibnamefont {Semmlinger}}, \bibinfo {author} {\bibfnamefont {Ming~Lun}\ \bibnamefont {Tseng}}, \bibinfo {author} {\bibfnamefont {Jian}\ \bibnamefont {Yang}}, \bibinfo {author} {\bibfnamefont {Ming}\ \bibnamefont {Zhang}}, \bibinfo {author} {\bibfnamefont {Chao}\ \bibnamefont {Zhang}}, \bibinfo {author} {\bibfnamefont {Wei-Yi}\ \bibnamefont {Tsai}}, \bibinfo {author} {\bibfnamefont {Din~Ping}\ \bibnamefont {Tsai}}, \bibinfo {author} {\bibfnamefont {Peter}\ \bibnamefont {Nordlander}}, \ and\ \bibinfo {author} {\bibfnamefont {Naomi~J.}\ \bibnamefont {Halas}},\ }\bibfield  {title} {\enquote {\bibinfo {title} {Vacuum ultraviolet light-generating metasurface},}\ }\href {\doibase 10.1021/acs.nanolett.8b01904} {\bibfield  {journal} {\bibinfo  {journal} {Nano Lett.}\ }\textbf {\bibinfo {volume} {18}},\ \bibinfo {pages} {5738--5743} (\bibinfo {year} {2018})}\BibitemShut {NoStop}%
\bibitem [{\citenamefont {Wan}\ \emph {et~al.}(2017)\citenamefont {Wan}, \citenamefont {Gao},\ and\ \citenamefont {Yang}}]{wan2017metasurface}%
  \BibitemOpen
  \bibfield  {author} {\bibinfo {author} {\bibfnamefont {Weiwei}\ \bibnamefont {Wan}}, \bibinfo {author} {\bibfnamefont {Jie}\ \bibnamefont {Gao}}, \ and\ \bibinfo {author} {\bibfnamefont {Xiaodong}\ \bibnamefont {Yang}},\ }\bibfield  {title} {\enquote {\bibinfo {title} {Metasurface holograms for holographic imaging},}\ }\href {\doibase 10.1002/adom.201700541} {\bibfield  {journal} {\bibinfo  {journal} {Adv.\ Opt.\ Mater.}\ }\textbf {\bibinfo {volume} {5}},\ \bibinfo {pages} {1700541} (\bibinfo {year} {2017})}\BibitemShut {NoStop}%
\bibitem [{\citenamefont {Deng}\ and\ \citenamefont {Li}(2017)}]{deng2017metasurface}%
  \BibitemOpen
  \bibfield  {author} {\bibinfo {author} {\bibfnamefont {Zi-Lan}\ \bibnamefont {Deng}}\ and\ \bibinfo {author} {\bibfnamefont {Guixin}\ \bibnamefont {Li}},\ }\bibfield  {title} {\enquote {\bibinfo {title} {Metasurface optical holography},}\ }\href {\doibase 10.1016/j.mtphys.2017.10.002} {\bibfield  {journal} {\bibinfo  {journal} {Mater.\ Today Phys.}\ }\textbf {\bibinfo {volume} {3}},\ \bibinfo {pages} {16--32} (\bibinfo {year} {2017})}\BibitemShut {NoStop}%
\bibitem [{\citenamefont {Karimi}\ \emph {et~al.}(2014)\citenamefont {Karimi}, \citenamefont {Schulz}, \citenamefont {Leon}, \citenamefont {Qassim}, \citenamefont {Upham},\ and\ \citenamefont {Boyd}}]{karimi2014generating}%
  \BibitemOpen
  \bibfield  {author} {\bibinfo {author} {\bibfnamefont {Ebrahim}\ \bibnamefont {Karimi}}, \bibinfo {author} {\bibfnamefont {Sebastian~A.}\ \bibnamefont {Schulz}}, \bibinfo {author} {\bibfnamefont {Israel~De}\ \bibnamefont {Leon}}, \bibinfo {author} {\bibfnamefont {Hammam}\ \bibnamefont {Qassim}}, \bibinfo {author} {\bibfnamefont {Jeremy}\ \bibnamefont {Upham}}, \ and\ \bibinfo {author} {\bibfnamefont {Robert~W.}\ \bibnamefont {Boyd}},\ }\bibfield  {title} {\enquote {\bibinfo {title} {Generating optical orbital angular momentum at visible wavelengths using a plasmonic metasurface},}\ }\href {\doibase 10.1038/lsa.2014.48} {\bibfield  {journal} {\bibinfo  {journal} {Light: Sci.\ Appl.}\ }\textbf {\bibinfo {volume} {3}},\ \bibinfo {pages} {e167} (\bibinfo {year} {2014})}\BibitemShut {NoStop}%
\bibitem [{\citenamefont {Wang}\ \emph {et~al.}(2018)\citenamefont {Wang}, \citenamefont {Wu}, \citenamefont {Su}, \citenamefont {Lai}, \citenamefont {Chen}, \citenamefont {Kuo}, \citenamefont {Chen}, \citenamefont {Chen}, \citenamefont {Huang}, \citenamefont {Wang} \emph {et~al.}}]{wang2018broadband}%
  \BibitemOpen
  \bibfield  {author} {\bibinfo {author} {\bibfnamefont {Shuming}\ \bibnamefont {Wang}}, \bibinfo {author} {\bibfnamefont {Pin~Chieh}\ \bibnamefont {Wu}}, \bibinfo {author} {\bibfnamefont {Vin-Cent}\ \bibnamefont {Su}}, \bibinfo {author} {\bibfnamefont {Yi-Chieh}\ \bibnamefont {Lai}}, \bibinfo {author} {\bibfnamefont {Mu-Ku}\ \bibnamefont {Chen}}, \bibinfo {author} {\bibfnamefont {Hsin~Yu}\ \bibnamefont {Kuo}}, \bibinfo {author} {\bibfnamefont {Bo~Han}\ \bibnamefont {Chen}}, \bibinfo {author} {\bibfnamefont {Yu~Han}\ \bibnamefont {Chen}}, \bibinfo {author} {\bibfnamefont {Tzu-Ting}\ \bibnamefont {Huang}}, \bibinfo {author} {\bibfnamefont {Jung-Hsi}\ \bibnamefont {Wang}},  \emph {et~al.},\ }\bibfield  {title} {\enquote {\bibinfo {title} {A broadband achromatic metalens in the visible},}\ }\href {\doibase 10.1038/s41565-017-0052-4} {\bibfield  {journal} {\bibinfo  {journal} {Nat.\ Nanotechnol.}\ }\textbf {\bibinfo {volume} {13}},\ \bibinfo {pages} {227--232} (\bibinfo {year} {2018})}\BibitemShut {NoStop}%
\bibitem [{\citenamefont {Deshpande}\ \emph {et~al.}(2018)\citenamefont {Deshpande}, \citenamefont {Zenin}, \citenamefont {Ding}, \citenamefont {Mortensen},\ and\ \citenamefont {Bozhevolnyi}}]{deshpande2018direct}%
  \BibitemOpen
  \bibfield  {author} {\bibinfo {author} {\bibfnamefont {Rucha}\ \bibnamefont {Deshpande}}, \bibinfo {author} {\bibfnamefont {Vladimir~A.}\ \bibnamefont {Zenin}}, \bibinfo {author} {\bibfnamefont {Fei}\ \bibnamefont {Ding}}, \bibinfo {author} {\bibfnamefont {N.~Asger}\ \bibnamefont {Mortensen}}, \ and\ \bibinfo {author} {\bibfnamefont {Sergey~I.}\ \bibnamefont {Bozhevolnyi}},\ }\bibfield  {title} {\enquote {\bibinfo {title} {Direct characterization of near-field coupling in gap plasmon-based metasurfaces},}\ }\href {\doibase 10.1021/acs.nanolett.8b03058} {\bibfield  {journal} {\bibinfo  {journal} {Nano Lett.}\ }\textbf {\bibinfo {volume} {18}},\ \bibinfo {pages} {6265--6270} (\bibinfo {year} {2018})}\BibitemShut {NoStop}%
\bibitem [{\citenamefont {Li}\ \emph {et~al.}(2020)\citenamefont {Li}, \citenamefont {Hu}, \citenamefont {Dolado}, \citenamefont {Tymchenko}, \citenamefont {Qiu}, \citenamefont {Alfaro-Mozaz}, \citenamefont {Casanova}, \citenamefont {Hueso}, \citenamefont {Liu}, \citenamefont {Edgar} \emph {et~al.}}]{li2020collective}%
  \BibitemOpen
  \bibfield  {author} {\bibinfo {author} {\bibfnamefont {Peining}\ \bibnamefont {Li}}, \bibinfo {author} {\bibfnamefont {Guangwei}\ \bibnamefont {Hu}}, \bibinfo {author} {\bibfnamefont {Irene}\ \bibnamefont {Dolado}}, \bibinfo {author} {\bibfnamefont {Mykhailo}\ \bibnamefont {Tymchenko}}, \bibinfo {author} {\bibfnamefont {Cheng-Wei}\ \bibnamefont {Qiu}}, \bibinfo {author} {\bibfnamefont {Francisco~Javier}\ \bibnamefont {Alfaro-Mozaz}}, \bibinfo {author} {\bibfnamefont {F{\`e}lix}\ \bibnamefont {Casanova}}, \bibinfo {author} {\bibfnamefont {Luis~E.}\ \bibnamefont {Hueso}}, \bibinfo {author} {\bibfnamefont {Song}\ \bibnamefont {Liu}}, \bibinfo {author} {\bibfnamefont {James~H.}\ \bibnamefont {Edgar}},  \emph {et~al.},\ }\bibfield  {title} {\enquote {\bibinfo {title} {Collective near-field coupling and nonlocal phenomena in infrared-phononic metasurfaces for nano-light canalization},}\ }\href {\doibase 10.1038/s41467-020-17435-x} {\bibfield  {journal} {\bibinfo  {journal} {Nat.\ Commun.}\ }\textbf {\bibinfo
  {volume} {11}},\ \bibinfo {pages} {3663} (\bibinfo {year} {2020})}\BibitemShut {NoStop}%
\bibitem [{\citenamefont {{Garc\'{\i}a de Abajo}}(2010{\natexlab{a}})}]{garciadeabajo2010optical}%
  \BibitemOpen
  \bibfield  {author} {\bibinfo {author} {\bibfnamefont {F.~Javier}\ \bibnamefont {{Garc\'{\i}a de Abajo}}},\ }\bibfield  {title} {\enquote {\bibinfo {title} {Optical excitations in electron microscopy},}\ }\href {\doibase 10.1103/RevModPhys.82.209} {\bibfield  {journal} {\bibinfo  {journal} {Rev.\ Mod.\ Phys.}\ }\textbf {\bibinfo {volume} {82}},\ \bibinfo {pages} {209} (\bibinfo {year} {2010}{\natexlab{a}})}\BibitemShut {NoStop}%
\bibitem [{\citenamefont {Adamo}\ \emph {et~al.}(2009)\citenamefont {Adamo}, \citenamefont {MacDonald}, \citenamefont {Fu}, \citenamefont {Wang}, \citenamefont {Tsai}, \citenamefont {{Garc\'{\i}a de Abajo}},\ and\ \citenamefont {Zheludev}}]{adamo2009light}%
  \BibitemOpen
  \bibfield  {author} {\bibinfo {author} {\bibfnamefont {Giorgio}\ \bibnamefont {Adamo}}, \bibinfo {author} {\bibfnamefont {Kevin~F.}\ \bibnamefont {MacDonald}}, \bibinfo {author} {\bibfnamefont {Y.~H.}\ \bibnamefont {Fu}}, \bibinfo {author} {\bibfnamefont {C.~M.}\ \bibnamefont {Wang}}, \bibinfo {author} {\bibfnamefont {Din~Ping}\ \bibnamefont {Tsai}}, \bibinfo {author} {\bibfnamefont {F.~Javier}\ \bibnamefont {{Garc\'{\i}a de Abajo}}}, \ and\ \bibinfo {author} {\bibfnamefont {N.~I.}\ \bibnamefont {Zheludev}},\ }\bibfield  {title} {\enquote {\bibinfo {title} {Light well: a tunable free-electron light source on a chip},}\ }\href {\doibase 10.1103/PhysRevLett.103.113901} {\bibfield  {journal} {\bibinfo  {journal} {Phys.\ Rev.\ Lett.}\ }\textbf {\bibinfo {volume} {103}},\ \bibinfo {pages} {113901} (\bibinfo {year} {2009})}\BibitemShut {NoStop}%
\bibitem [{\citenamefont {Rosolen}\ \emph {et~al.}(2018)\citenamefont {Rosolen}, \citenamefont {Wong}, \citenamefont {Rivera}, \citenamefont {Maes}, \citenamefont {Solja{\v{c}}i{\'c}},\ and\ \citenamefont {Kaminer}}]{rosolen2018metasurface}%
  \BibitemOpen
  \bibfield  {author} {\bibinfo {author} {\bibfnamefont {Gilles}\ \bibnamefont {Rosolen}}, \bibinfo {author} {\bibfnamefont {Liang~Jie}\ \bibnamefont {Wong}}, \bibinfo {author} {\bibfnamefont {Nicholas}\ \bibnamefont {Rivera}}, \bibinfo {author} {\bibfnamefont {Bjorn}\ \bibnamefont {Maes}}, \bibinfo {author} {\bibfnamefont {Marin}\ \bibnamefont {Solja{\v{c}}i{\'c}}}, \ and\ \bibinfo {author} {\bibfnamefont {Ido}\ \bibnamefont {Kaminer}},\ }\bibfield  {title} {\enquote {\bibinfo {title} {Metasurface-based multi-harmonic free-electron light source},}\ }\href {\doibase 10.1038/s41377-018-0068-x} {\bibfield  {journal} {\bibinfo  {journal} {Light: Sci.\ Appl.}\ }\textbf {\bibinfo {volume} {7}},\ \bibinfo {pages} {64} (\bibinfo {year} {2018})}\BibitemShut {NoStop}%
\bibitem [{\citenamefont {Smith}\ and\ \citenamefont {Purcell}(1953)}]{smith1953visible}%
  \BibitemOpen
  \bibfield  {author} {\bibinfo {author} {\bibfnamefont {Stephen~J.}\ \bibnamefont {Smith}}\ and\ \bibinfo {author} {\bibfnamefont {Edward~M.}\ \bibnamefont {Purcell}},\ }\bibfield  {title} {\enquote {\bibinfo {title} {Visible light from localized surface charges moving across a grating},}\ }\href {\doibase 10.1103/PhysRev.92.1069} {\bibfield  {journal} {\bibinfo  {journal} {Phys.\ Rev.}\ }\textbf {\bibinfo {volume} {92}},\ \bibinfo {pages} {1069} (\bibinfo {year} {1953})}\BibitemShut {NoStop}%
\bibitem [{\citenamefont {Su}\ \emph {et~al.}(2019)\citenamefont {Su}, \citenamefont {Xiong}, \citenamefont {Xu}, \citenamefont {Cai}, \citenamefont {Yin}, \citenamefont {Peng},\ and\ \citenamefont {Liu}}]{su2019manipulating}%
  \BibitemOpen
  \bibfield  {author} {\bibinfo {author} {\bibfnamefont {Zhaoxian}\ \bibnamefont {Su}}, \bibinfo {author} {\bibfnamefont {Bo}~\bibnamefont {Xiong}}, \bibinfo {author} {\bibfnamefont {Yihao}\ \bibnamefont {Xu}}, \bibinfo {author} {\bibfnamefont {Ziqiang}\ \bibnamefont {Cai}}, \bibinfo {author} {\bibfnamefont {Jianbo}\ \bibnamefont {Yin}}, \bibinfo {author} {\bibfnamefont {Ruwen}\ \bibnamefont {Peng}}, \ and\ \bibinfo {author} {\bibfnamefont {Yongmin}\ \bibnamefont {Liu}},\ }\bibfield  {title} {\enquote {\bibinfo {title} {Manipulating cherenkov radiation and smith--purcell radiation by artificial structures},}\ }\href {\doibase 10.1002/adom.201801666} {\bibfield  {journal} {\bibinfo  {journal} {Adv.\ Opt.\ Mater.}\ }\textbf {\bibinfo {volume} {7}},\ \bibinfo {pages} {1801666} (\bibinfo {year} {2019})}\BibitemShut {NoStop}%
\bibitem [{\citenamefont {{Garc\'{\i}a de Abajo}}\ and\ \citenamefont {Howie}(1998)}]{paper014}%
  \BibitemOpen
  \bibfield  {author} {\bibinfo {author} {\bibfnamefont {F.~J.}\ \bibnamefont {{Garc\'{\i}a de Abajo}}}\ and\ \bibinfo {author} {\bibfnamefont {A.}~\bibnamefont {Howie}},\ }\bibfield  {title} {\enquote {\bibinfo {title} {Relativistic electron energy loss and electron-induced photon emission in inhomogeneous dielectrics},}\ }\href {\doibase 10.1103/PhysRevLett.80.5180} {\bibfield  {journal} {\bibinfo  {journal} {Phys.\ Rev.\ Lett.}\ }\textbf {\bibinfo {volume} {80}},\ \bibinfo {pages} {5180--5183} (\bibinfo {year} {1998})}\BibitemShut {NoStop}%
\bibitem [{\citenamefont {Wang}\ \emph {et~al.}(2016)\citenamefont {Wang}, \citenamefont {Yao}, \citenamefont {Chen}, \citenamefont {Chen},\ and\ \citenamefont {Liu}}]{wang2016manipulating}%
  \BibitemOpen
  \bibfield  {author} {\bibinfo {author} {\bibfnamefont {Zuojia}\ \bibnamefont {Wang}}, \bibinfo {author} {\bibfnamefont {Kan}\ \bibnamefont {Yao}}, \bibinfo {author} {\bibfnamefont {Min}\ \bibnamefont {Chen}}, \bibinfo {author} {\bibfnamefont {Hongsheng}\ \bibnamefont {Chen}}, \ and\ \bibinfo {author} {\bibfnamefont {Yongmin}\ \bibnamefont {Liu}},\ }\bibfield  {title} {\enquote {\bibinfo {title} {Manipulating smith-purcell emission with babinet metasurfaces},}\ }\href {\doibase 10.1103/PhysRevLett.117.157401} {\bibfield  {journal} {\bibinfo  {journal} {Phys.\ Rev.\ Lett.}\ }\textbf {\bibinfo {volume} {117}},\ \bibinfo {pages} {157401} (\bibinfo {year} {2016})}\BibitemShut {NoStop}%
\bibitem [{\citenamefont {Garaev}\ \emph {et~al.}(2021)\citenamefont {Garaev}, \citenamefont {Sergeeva},\ and\ \citenamefont {Tishchenko}}]{garaev2021theory}%
  \BibitemOpen
  \bibfield  {author} {\bibinfo {author} {\bibfnamefont {D.~I.}\ \bibnamefont {Garaev}}, \bibinfo {author} {\bibfnamefont {D.~Yu.}\ \bibnamefont {Sergeeva}}, \ and\ \bibinfo {author} {\bibfnamefont {A.~A.}\ \bibnamefont {Tishchenko}},\ }\bibfield  {title} {\enquote {\bibinfo {title} {Theory of smith-purcell radiation from a 2d array of small noninteracting particles},}\ }\href {\doibase 10.1103/PhysRevB.103.075403} {\bibfield  {journal} {\bibinfo  {journal} {Phys.\ Rev.\ B}\ }\textbf {\bibinfo {volume} {103}},\ \bibinfo {pages} {075403} (\bibinfo {year} {2021})}\BibitemShut {NoStop}%
\bibitem [{\citenamefont {Karnieli}\ \emph {et~al.}(2022)\citenamefont {Karnieli}, \citenamefont {Roitman}, \citenamefont {Liebtrau}, \citenamefont {Tsesses}, \citenamefont {Nielen}, \citenamefont {Kaminer}, \citenamefont {Arie},\ and\ \citenamefont {Polman}}]{karnieli2022cylindrical}%
  \BibitemOpen
  \bibfield  {author} {\bibinfo {author} {\bibfnamefont {Aviv}\ \bibnamefont {Karnieli}}, \bibinfo {author} {\bibfnamefont {Dolev}\ \bibnamefont {Roitman}}, \bibinfo {author} {\bibfnamefont {Matthias}\ \bibnamefont {Liebtrau}}, \bibinfo {author} {\bibfnamefont {Shai}\ \bibnamefont {Tsesses}}, \bibinfo {author} {\bibfnamefont {Nika~Van}\ \bibnamefont {Nielen}}, \bibinfo {author} {\bibfnamefont {Ido}\ \bibnamefont {Kaminer}}, \bibinfo {author} {\bibfnamefont {Ady}\ \bibnamefont {Arie}}, \ and\ \bibinfo {author} {\bibfnamefont {Albert}\ \bibnamefont {Polman}},\ }\bibfield  {title} {\enquote {\bibinfo {title} {Cylindrical metalens for generation and focusing of free-electron radiation},}\ }\href {\doibase 10.1021/acs.nanolett.2c01416} {\bibfield  {journal} {\bibinfo  {journal} {Nano Lett.}\ }\textbf {\bibinfo {volume} {22}},\ \bibinfo {pages} {5641--5650} (\bibinfo {year} {2022})}\BibitemShut {NoStop}%
\bibitem [{\citenamefont {Saavedra}\ \emph {et~al.}(2016)\citenamefont {Saavedra}, \citenamefont {Castells-Graells},\ and\ \citenamefont {{Garc\'{\i}a de Abajo}}}]{saavedra2016smith}%
  \BibitemOpen
  \bibfield  {author} {\bibinfo {author} {\bibfnamefont {J.~R.~M.}\ \bibnamefont {Saavedra}}, \bibinfo {author} {\bibfnamefont {D.}~\bibnamefont {Castells-Graells}}, \ and\ \bibinfo {author} {\bibfnamefont {F.~Javier}\ \bibnamefont {{Garc\'{\i}a de Abajo}}},\ }\bibfield  {title} {\enquote {\bibinfo {title} {Smith-purcell radiation emission in aperiodic arrays},}\ }\href {\doibase 10.1103/PhysRevB.94.035418} {\bibfield  {journal} {\bibinfo  {journal} {Phys.\ Rev.\ B}\ }\textbf {\bibinfo {volume} {94}},\ \bibinfo {pages} {035418} (\bibinfo {year} {2016})}\BibitemShut {NoStop}%
\bibitem [{\citenamefont {Cueff}\ \emph {et~al.}(2020)\citenamefont {Cueff}, \citenamefont {Gentile}, \citenamefont {Lauret} \emph {et~al.}}]{CGL20}%
  \BibitemOpen
  \bibfield  {author} {\bibinfo {author} {\bibfnamefont {S{\'e}bastien}\ \bibnamefont {Cueff}}, \bibinfo {author} {\bibfnamefont {Mario~J.}\ \bibnamefont {Gentile}}, \bibinfo {author} {\bibfnamefont {Jean-S{\'e}bastien}\ \bibnamefont {Lauret}},  \emph {et~al.},\ }\bibfield  {title} {\enquote {\bibinfo {title} {Vo\(_2\) nanophotonics},}\ }\href {\doibase 10.1063/1.5143815} {\bibfield  {journal} {\bibinfo  {journal} {APL Photonics}\ }\textbf {\bibinfo {volume} {5}},\ \bibinfo {pages} {110901} (\bibinfo {year} {2020})}\BibitemShut {NoStop}%
\bibitem [{\citenamefont {Driscoll}\ \emph {et~al.}(2008)\citenamefont {Driscoll}, \citenamefont {Palit}, \citenamefont {Qazilbash} \emph {et~al.}}]{DPQ08}%
  \BibitemOpen
  \bibfield  {author} {\bibinfo {author} {\bibfnamefont {T.}~\bibnamefont {Driscoll}}, \bibinfo {author} {\bibfnamefont {S.}~\bibnamefont {Palit}}, \bibinfo {author} {\bibfnamefont {M.~M.}\ \bibnamefont {Qazilbash}},  \emph {et~al.},\ }\bibfield  {title} {\enquote {\bibinfo {title} {Dynamic tuning of an infrared hybrid-metamaterial resonance using vanadium dioxide},}\ }\href {\doibase 10.1063/1.2956675} {\bibfield  {journal} {\bibinfo  {journal} {Applied Physics Letters}\ }\textbf {\bibinfo {volume} {93}},\ \bibinfo {pages} {024101} (\bibinfo {year} {2008})}\BibitemShut {NoStop}%
\bibitem [{\citenamefont {Verleur}\ \emph {et~al.}(1968)\citenamefont {Verleur}, \citenamefont {Barker},\ and\ \citenamefont {Berglund}}]{VBB68}%
  \BibitemOpen
  \bibfield  {author} {\bibinfo {author} {\bibfnamefont {H.~W.}\ \bibnamefont {Verleur}}, \bibinfo {author} {\bibfnamefont {Jr.}\ \bibnamefont {Barker}, \bibfnamefont {A.~S.}}, \ and\ \bibinfo {author} {\bibfnamefont {C.~N.}\ \bibnamefont {Berglund}},\ }\bibfield  {title} {\enquote {\bibinfo {title} {Optical properties of vo$_2$ between 0.25 and 5 ev},}\ }\href {\doibase 10.1103/PhysRev.172.788} {\bibfield  {journal} {\bibinfo  {journal} {Physical Review}\ }\textbf {\bibinfo {volume} {172}},\ \bibinfo {pages} {788--798} (\bibinfo {year} {1968})}\BibitemShut {NoStop}%
\bibitem [{\citenamefont {Qazilbash}\ \emph {et~al.}(2007)\citenamefont {Qazilbash}, \citenamefont {Brehm}, \citenamefont {Chae}, \citenamefont {Ho}, \citenamefont {Andreev}, \citenamefont {Kim}, \citenamefont {Yun}, \citenamefont {Balatsky}, \citenamefont {Maple}, \citenamefont {Keilmann}, \citenamefont {Kim},\ and\ \citenamefont {Basov}}]{QBC07}%
  \BibitemOpen
  \bibfield  {author} {\bibinfo {author} {\bibfnamefont {M.~M.}\ \bibnamefont {Qazilbash}}, \bibinfo {author} {\bibfnamefont {M.}~\bibnamefont {Brehm}}, \bibinfo {author} {\bibfnamefont {Byung-Gyu}\ \bibnamefont {Chae}}, \bibinfo {author} {\bibfnamefont {P.-C.}\ \bibnamefont {Ho}}, \bibinfo {author} {\bibfnamefont {G.~O.}\ \bibnamefont {Andreev}}, \bibinfo {author} {\bibfnamefont {Bong-Jun}\ \bibnamefont {Kim}}, \bibinfo {author} {\bibfnamefont {Sun~Jin}\ \bibnamefont {Yun}}, \bibinfo {author} {\bibfnamefont {A.~V.}\ \bibnamefont {Balatsky}}, \bibinfo {author} {\bibfnamefont {M.~B.}\ \bibnamefont {Maple}}, \bibinfo {author} {\bibfnamefont {F.}~\bibnamefont {Keilmann}}, \bibinfo {author} {\bibfnamefont {Hyun-Tak}\ \bibnamefont {Kim}}, \ and\ \bibinfo {author} {\bibfnamefont {D.~N.}\ \bibnamefont {Basov}},\ }\bibfield  {title} {\enquote {\bibinfo {title} {Mott transition in vo$_2$ revealed by infrared spectroscopy and nano-imaging},}\ }\href {\doibase 10.1126/science.1150124} {\bibfield  {journal} {\bibinfo
  {journal} {Science}\ }\textbf {\bibinfo {volume} {318}},\ \bibinfo {pages} {1750--1753} (\bibinfo {year} {2007})}\BibitemShut {NoStop}%
\bibitem [{\citenamefont {Wen}\ \emph {et~al.}(2013)\citenamefont {Wen}, \citenamefont {Guo}, \citenamefont {Barnes}, \citenamefont {Lee}, \citenamefont {Walko}, \citenamefont {Schaller}, \citenamefont {Moyer}, \citenamefont {Misra}, \citenamefont {Li}, \citenamefont {Dufresne}, \citenamefont {Schlom}, \citenamefont {Gopalan},\ and\ \citenamefont {Freeland}}]{WGB13}%
  \BibitemOpen
  \bibfield  {author} {\bibinfo {author} {\bibfnamefont {Haidan}\ \bibnamefont {Wen}}, \bibinfo {author} {\bibfnamefont {Lu}~\bibnamefont {Guo}}, \bibinfo {author} {\bibfnamefont {Eftihia}\ \bibnamefont {Barnes}}, \bibinfo {author} {\bibfnamefont {June~Hyuk}\ \bibnamefont {Lee}}, \bibinfo {author} {\bibfnamefont {Donald~A.}\ \bibnamefont {Walko}}, \bibinfo {author} {\bibfnamefont {Richard~D.}\ \bibnamefont {Schaller}}, \bibinfo {author} {\bibfnamefont {Jarrett~A.}\ \bibnamefont {Moyer}}, \bibinfo {author} {\bibfnamefont {Rajiv}\ \bibnamefont {Misra}}, \bibinfo {author} {\bibfnamefont {Yuelin}\ \bibnamefont {Li}}, \bibinfo {author} {\bibfnamefont {Eric~M.}\ \bibnamefont {Dufresne}}, \bibinfo {author} {\bibfnamefont {Darrell~G.}\ \bibnamefont {Schlom}}, \bibinfo {author} {\bibfnamefont {Venkatraman}\ \bibnamefont {Gopalan}}, \ and\ \bibinfo {author} {\bibfnamefont {John~W.}\ \bibnamefont {Freeland}},\ }\bibfield  {title} {\enquote {\bibinfo {title} {Structural and electronic recovery pathways of a photoexcited
  ultrathin vo$_2$ film},}\ }\href {\doibase 10.1103/PhysRevB.88.165424} {\bibfield  {journal} {\bibinfo  {journal} {Physical Review B}\ }\textbf {\bibinfo {volume} {88}},\ \bibinfo {pages} {165424} (\bibinfo {year} {2013})}\BibitemShut {NoStop}%
\bibitem [{\citenamefont {Lysenko}\ \emph {et~al.}(2017)\citenamefont {Lysenko}, \citenamefont {Kumar}, \citenamefont {R{\'u}a}, \citenamefont {Figueroa}, \citenamefont {Fernandez},\ and\ \citenamefont {Liu}}]{LKR17}%
  \BibitemOpen
  \bibfield  {author} {\bibinfo {author} {\bibfnamefont {Sergiy}\ \bibnamefont {Lysenko}}, \bibinfo {author} {\bibfnamefont {Nardeep}\ \bibnamefont {Kumar}}, \bibinfo {author} {\bibfnamefont {Armando}\ \bibnamefont {R{\'u}a}}, \bibinfo {author} {\bibfnamefont {Jos{\'e}}\ \bibnamefont {Figueroa}}, \bibinfo {author} {\bibfnamefont {Felix}\ \bibnamefont {Fernandez}}, \ and\ \bibinfo {author} {\bibfnamefont {Hua}\ \bibnamefont {Liu}},\ }\bibfield  {title} {\enquote {\bibinfo {title} {Ultrafast structural dynamics of vo$_2$},}\ }\href {\doibase 10.1103/PhysRevB.96.075128} {\bibfield  {journal} {\bibinfo  {journal} {Physical Review B}\ }\textbf {\bibinfo {volume} {96}},\ \bibinfo {pages} {075128} (\bibinfo {year} {2017})}\BibitemShut {NoStop}%
\bibitem [{\citenamefont {Muskens}\ \emph {et~al.}(2016)\citenamefont {Muskens}, \citenamefont {Bergamini}, \citenamefont {Wang}, \citenamefont {Gaskell}, \citenamefont {Zabala}, \citenamefont {de~Groot}, \citenamefont {Sheel},\ and\ \citenamefont {Aizpurua}}]{MBW16}%
  \BibitemOpen
  \bibfield  {author} {\bibinfo {author} {\bibfnamefont {Otto~L.}\ \bibnamefont {Muskens}}, \bibinfo {author} {\bibfnamefont {Luca}\ \bibnamefont {Bergamini}}, \bibinfo {author} {\bibfnamefont {Yudong}\ \bibnamefont {Wang}}, \bibinfo {author} {\bibfnamefont {Jeffrey~M.}\ \bibnamefont {Gaskell}}, \bibinfo {author} {\bibfnamefont {Nerea}\ \bibnamefont {Zabala}}, \bibinfo {author} {\bibfnamefont {C.~H.}\ \bibnamefont {de~Groot}}, \bibinfo {author} {\bibfnamefont {David~W.}\ \bibnamefont {Sheel}}, \ and\ \bibinfo {author} {\bibfnamefont {Javier}\ \bibnamefont {Aizpurua}},\ }\bibfield  {title} {\enquote {\bibinfo {title} {Antenna-assisted picosecond control of nanoscale phase transition in vanadium dioxide},}\ }\href {\doibase 10.1038/lsa.2016.173} {\bibfield  {journal} {\bibinfo  {journal} {Light: Science \& Applications}\ }\textbf {\bibinfo {volume} {5}},\ \bibinfo {pages} {e16173} (\bibinfo {year} {2016})}\BibitemShut {NoStop}%
\bibitem [{\citenamefont {Martens}\ \emph {et~al.}(2014)\citenamefont {Martens}, \citenamefont {Aetukuri}, \citenamefont {Jeong}, \citenamefont {Samant},\ and\ \citenamefont {Parkin}}]{MAJ14}%
  \BibitemOpen
  \bibfield  {author} {\bibinfo {author} {\bibfnamefont {Koen}\ \bibnamefont {Martens}}, \bibinfo {author} {\bibfnamefont {Nagaphani}\ \bibnamefont {Aetukuri}}, \bibinfo {author} {\bibfnamefont {Jaewoo}\ \bibnamefont {Jeong}}, \bibinfo {author} {\bibfnamefont {Mahesh~G.}\ \bibnamefont {Samant}}, \ and\ \bibinfo {author} {\bibfnamefont {Stuart S.~P.}\ \bibnamefont {Parkin}},\ }\bibfield  {title} {\enquote {\bibinfo {title} {Improved metal-insulator-transition characteristics of ultrathin vo$_2$ epitaxial films by optimized surface preparation of rutile tio$_2$ substrates},}\ }\href {\doibase 10.1063/1.4866037} {\bibfield  {journal} {\bibinfo  {journal} {Applied Physics Letters}\ }\textbf {\bibinfo {volume} {104}},\ \bibinfo {pages} {081918} (\bibinfo {year} {2014})}\BibitemShut {NoStop}%
\bibitem [{\citenamefont {Jepsen}\ \emph {et~al.}(2006)\citenamefont {Jepsen}, \citenamefont {Fischer}, \citenamefont {Thoman}, \citenamefont {Helm}, \citenamefont {Suh}, \citenamefont {Lopez},\ and\ \citenamefont {Haglund}}]{JFT06}%
  \BibitemOpen
  \bibfield  {author} {\bibinfo {author} {\bibfnamefont {Peter~Uhd}\ \bibnamefont {Jepsen}}, \bibinfo {author} {\bibfnamefont {Bernd~M.}\ \bibnamefont {Fischer}}, \bibinfo {author} {\bibfnamefont {Andreas}\ \bibnamefont {Thoman}}, \bibinfo {author} {\bibfnamefont {Hanspeter}\ \bibnamefont {Helm}}, \bibinfo {author} {\bibfnamefont {J.~Y.}\ \bibnamefont {Suh}}, \bibinfo {author} {\bibfnamefont {Ren{\'e}}\ \bibnamefont {Lopez}}, \ and\ \bibinfo {author} {\bibfnamefont {Jr.}\ \bibnamefont {Haglund}, \bibfnamefont {R.~F.}},\ }\bibfield  {title} {\enquote {\bibinfo {title} {Metal-insulator phase transition in a vo$_2$ thin film observed with terahertz spectroscopy},}\ }\href {\doibase 10.1103/PhysRevB.74.205103} {\bibfield  {journal} {\bibinfo  {journal} {Physical Review B}\ }\textbf {\bibinfo {volume} {74}},\ \bibinfo {pages} {205103} (\bibinfo {year} {2006})}\BibitemShut {NoStop}%
\bibitem [{\citenamefont {Pashkin}\ \emph {et~al.}(2011)\citenamefont {Pashkin}, \citenamefont {K{\"u}bler}, \citenamefont {Ehrke}, \citenamefont {Lopez}, \citenamefont {Halabica}, \citenamefont {Haglund}, \citenamefont {Huber},\ and\ \citenamefont {Leitenstorfer}}]{PKE11}%
  \BibitemOpen
  \bibfield  {author} {\bibinfo {author} {\bibfnamefont {A.}~\bibnamefont {Pashkin}}, \bibinfo {author} {\bibfnamefont {C.}~\bibnamefont {K{\"u}bler}}, \bibinfo {author} {\bibfnamefont {H.}~\bibnamefont {Ehrke}}, \bibinfo {author} {\bibfnamefont {R.}~\bibnamefont {Lopez}}, \bibinfo {author} {\bibfnamefont {A.}~\bibnamefont {Halabica}}, \bibinfo {author} {\bibfnamefont {Jr.}\ \bibnamefont {Haglund}, \bibfnamefont {R.~F.}}, \bibinfo {author} {\bibfnamefont {R.}~\bibnamefont {Huber}}, \ and\ \bibinfo {author} {\bibfnamefont {A.}~\bibnamefont {Leitenstorfer}},\ }\bibfield  {title} {\enquote {\bibinfo {title} {Ultrafast insulator-metal phase transition in vo$_2$ studied by multiterahertz spectroscopy},}\ }\href {\doibase 10.1103/PhysRevB.83.195120} {\bibfield  {journal} {\bibinfo  {journal} {Physical Review B}\ }\textbf {\bibinfo {volume} {83}},\ \bibinfo {pages} {195120} (\bibinfo {year} {2011})}\BibitemShut {NoStop}%
\bibitem [{\citenamefont {{Garc\'{\i}a de Abajo}}(2014)}]{paper235}%
  \BibitemOpen
  \bibfield  {author} {\bibinfo {author} {\bibfnamefont {F.~J.}\ \bibnamefont {{Garc\'{\i}a de Abajo}}},\ }\bibfield  {title} {\enquote {\bibinfo {title} {Graphene plasmonics: challenges and opportunities},}\ }\href {\doibase 10.1021/ph400147y} {\bibfield  {journal} {\bibinfo  {journal} {ACS\ Photonics}\ }\textbf {\bibinfo {volume} {1}},\ \bibinfo {pages} {135--152} (\bibinfo {year} {2014})}\BibitemShut {NoStop}%
\bibitem [{\citenamefont {Grigorenko}\ \emph {et~al.}(2012)\citenamefont {Grigorenko}, \citenamefont {Polini},\ and\ \citenamefont {Novoselov}}]{GPN12}%
  \BibitemOpen
  \bibfield  {author} {\bibinfo {author} {\bibfnamefont {A.~N.}\ \bibnamefont {Grigorenko}}, \bibinfo {author} {\bibfnamefont {M.}~\bibnamefont {Polini}}, \ and\ \bibinfo {author} {\bibfnamefont {K.~S.}\ \bibnamefont {Novoselov}},\ }\bibfield  {title} {\enquote {\bibinfo {title} {Graphene plasmonics},}\ }\href {\doibase 10.1038/NPHOTON.2012.262} {\bibfield  {journal} {\bibinfo  {journal} {Nat.\ Photonics}\ }\textbf {\bibinfo {volume} {6}},\ \bibinfo {pages} {749--758} (\bibinfo {year} {2012})}\BibitemShut {NoStop}%
\bibitem [{\citenamefont {Silveiro}\ \emph {et~al.}(2015)\citenamefont {Silveiro}, \citenamefont {{Plaza Ortega}},\ and\ \citenamefont {{Garc\'{\i}a de Abajo}}}]{paper257}%
  \BibitemOpen
  \bibfield  {author} {\bibinfo {author} {\bibfnamefont {I.}~\bibnamefont {Silveiro}}, \bibinfo {author} {\bibfnamefont {J.~M.}\ \bibnamefont {{Plaza Ortega}}}, \ and\ \bibinfo {author} {\bibfnamefont {F.~J.}\ \bibnamefont {{Garc\'{\i}a de Abajo}}},\ }\bibfield  {title} {\enquote {\bibinfo {title} {Plasmon wave function of graphene nanoribbons},}\ }\href {\doibase 10.1088/1367-2630/17/8/083013} {\bibfield  {journal} {\bibinfo  {journal} {New\ J.\ Phys.}\ }\textbf {\bibinfo {volume} {17}},\ \bibinfo {pages} {083013} (\bibinfo {year} {2015})}\BibitemShut {NoStop}%
\bibitem [{\citenamefont {Mkhitaryan}\ \emph {et~al.}(2021)\citenamefont {Mkhitaryan}, \citenamefont {Dias}, \citenamefont {Carbone},\ and\ \citenamefont {{Garc\'{\i}a de Abajo}}}]{paper364}%
  \BibitemOpen
  \bibfield  {author} {\bibinfo {author} {\bibfnamefont {V.}~\bibnamefont {Mkhitaryan}}, \bibinfo {author} {\bibfnamefont {E.~J.~C.}\ \bibnamefont {Dias}}, \bibinfo {author} {\bibfnamefont {F.}~\bibnamefont {Carbone}}, \ and\ \bibinfo {author} {\bibfnamefont {F.~J.}\ \bibnamefont {{Garc\'{\i}a de Abajo}}},\ }\bibfield  {title} {\enquote {\bibinfo {title} {Ultrafast momentum-resolved free-electron probing of optically pumped plasmon thermal dynamics},}\ }\href {\doibase 10.1021/acsphotonics.0c01758} {\bibfield  {journal} {\bibinfo  {journal} {ACS\ Photonics}\ }\textbf {\bibinfo {volume} {8}},\ \bibinfo {pages} {614--624} (\bibinfo {year} {2021})}\BibitemShut {NoStop}%
\bibitem [{\citenamefont {Jiao}\ \emph {et~al.}(2009)\citenamefont {Jiao}, \citenamefont {Zhang}, \citenamefont {Wang}, \citenamefont {Diankov},\ and\ \citenamefont {Dai}}]{JZW09}%
  \BibitemOpen
  \bibfield  {author} {\bibinfo {author} {\bibfnamefont {Liying}\ \bibnamefont {Jiao}}, \bibinfo {author} {\bibfnamefont {Li}~\bibnamefont {Zhang}}, \bibinfo {author} {\bibfnamefont {Xinran}\ \bibnamefont {Wang}}, \bibinfo {author} {\bibfnamefont {Georgi}\ \bibnamefont {Diankov}}, \ and\ \bibinfo {author} {\bibfnamefont {Hongjie}\ \bibnamefont {Dai}},\ }\bibfield  {title} {\enquote {\bibinfo {title} {Narrow graphene nanoribbons from carbon nanotubes},}\ }\href {\doibase 10.1038/nature07919} {\bibfield  {journal} {\bibinfo  {journal} {Nature}\ }\textbf {\bibinfo {volume} {458}},\ \bibinfo {pages} {877--880} (\bibinfo {year} {2009})}\BibitemShut {NoStop}%
\bibitem [{\citenamefont {Cai}\ \emph {et~al.}(2010)\citenamefont {Cai}, \citenamefont {Ruffieux}, \citenamefont {Jaafar}, \citenamefont {Bieri}, \citenamefont {Braun}, \citenamefont {Blankenburg}, \citenamefont {Muoth}, \citenamefont {Seitsonen}, \citenamefont {Saleh}, \citenamefont {Feng}, \citenamefont {M{\"u}llen},\ and\ \citenamefont {Fasel}}]{CRF10}%
  \BibitemOpen
  \bibfield  {author} {\bibinfo {author} {\bibfnamefont {Jinming}\ \bibnamefont {Cai}}, \bibinfo {author} {\bibfnamefont {Pascal}\ \bibnamefont {Ruffieux}}, \bibinfo {author} {\bibfnamefont {Rached}\ \bibnamefont {Jaafar}}, \bibinfo {author} {\bibfnamefont {Marco}\ \bibnamefont {Bieri}}, \bibinfo {author} {\bibfnamefont {Thomas}\ \bibnamefont {Braun}}, \bibinfo {author} {\bibfnamefont {Stephan}\ \bibnamefont {Blankenburg}}, \bibinfo {author} {\bibfnamefont {Matthias}\ \bibnamefont {Muoth}}, \bibinfo {author} {\bibfnamefont {Ari~P.}\ \bibnamefont {Seitsonen}}, \bibinfo {author} {\bibfnamefont {Moussa}\ \bibnamefont {Saleh}}, \bibinfo {author} {\bibfnamefont {Xinliang}\ \bibnamefont {Feng}}, \bibinfo {author} {\bibfnamefont {Klaus}\ \bibnamefont {M{\"u}llen}}, \ and\ \bibinfo {author} {\bibfnamefont {Roman}\ \bibnamefont {Fasel}},\ }\bibfield  {title} {\enquote {\bibinfo {title} {Atomically precise bottom-up fabrication of graphene nanoribbons},}\ }\href {\doibase 10.1038/nature09211} {\bibfield  {journal}
  {\bibinfo  {journal} {Nature}\ }\textbf {\bibinfo {volume} {466}},\ \bibinfo {pages} {470--473} (\bibinfo {year} {2010})}\BibitemShut {NoStop}%
\bibitem [{\citenamefont {Suk}\ \emph {et~al.}(2011)\citenamefont {Suk}, \citenamefont {Kitt}, \citenamefont {Magnuson}, \citenamefont {Hao}, \citenamefont {Ahmed}, \citenamefont {An}, \citenamefont {Swan}, \citenamefont {Goldberg},\ and\ \citenamefont {Ruoff}}]{SKM11}%
  \BibitemOpen
  \bibfield  {author} {\bibinfo {author} {\bibfnamefont {Ji~Won}\ \bibnamefont {Suk}}, \bibinfo {author} {\bibfnamefont {Alexander}\ \bibnamefont {Kitt}}, \bibinfo {author} {\bibfnamefont {Carl~W}\ \bibnamefont {Magnuson}}, \bibinfo {author} {\bibfnamefont {Yufeng}\ \bibnamefont {Hao}}, \bibinfo {author} {\bibfnamefont {Samir}\ \bibnamefont {Ahmed}}, \bibinfo {author} {\bibfnamefont {Jinho}\ \bibnamefont {An}}, \bibinfo {author} {\bibfnamefont {Anna~K}\ \bibnamefont {Swan}}, \bibinfo {author} {\bibfnamefont {Bennett~B}\ \bibnamefont {Goldberg}}, \ and\ \bibinfo {author} {\bibfnamefont {Rodney~S}\ \bibnamefont {Ruoff}},\ }\bibfield  {title} {\enquote {\bibinfo {title} {Transfer of cvd-grown monolayer graphene onto arbitrary substrates},}\ }\href@noop {} {\bibfield  {journal} {\bibinfo  {journal} {ACS nano}\ }\textbf {\bibinfo {volume} {5}},\ \bibinfo {pages} {6916--6924} (\bibinfo {year} {2011})}\BibitemShut {NoStop}%
\bibitem [{\citenamefont {Mutlu}\ \emph {et~al.}(2021)\citenamefont {Mutlu}, \citenamefont {Llin{\'a}s}, \citenamefont {Jacobse}, \citenamefont {Piskun}, \citenamefont {Blackwell}, \citenamefont {Crommie}, \citenamefont {Fischer},\ and\ \citenamefont {Bokor}}]{MLJ21}%
  \BibitemOpen
  \bibfield  {author} {\bibinfo {author} {\bibfnamefont {Zafer}\ \bibnamefont {Mutlu}}, \bibinfo {author} {\bibfnamefont {Juan~Pablo}\ \bibnamefont {Llin{\'a}s}}, \bibinfo {author} {\bibfnamefont {Peter~H.}\ \bibnamefont {Jacobse}}, \bibinfo {author} {\bibfnamefont {Ilya}\ \bibnamefont {Piskun}}, \bibinfo {author} {\bibfnamefont {Raymond}\ \bibnamefont {Blackwell}}, \bibinfo {author} {\bibfnamefont {Michael~F.}\ \bibnamefont {Crommie}}, \bibinfo {author} {\bibfnamefont {Felix~R.}\ \bibnamefont {Fischer}}, \ and\ \bibinfo {author} {\bibfnamefont {Jeffrey}\ \bibnamefont {Bokor}},\ }\bibfield  {title} {\enquote {\bibinfo {title} {Transfer-free synthesis of atomically precise graphene nanoribbons on insulating substrates},}\ }\href {\doibase 10.1021/acsnano.0c07591} {\bibfield  {journal} {\bibinfo  {journal} {ACS Nano}\ }\textbf {\bibinfo {volume} {15}},\ \bibinfo {pages} {2635--2642} (\bibinfo {year} {2021})}\BibitemShut {NoStop}%
\bibitem [{\citenamefont {Zhang}\ \emph {et~al.}(2023)\citenamefont {Zhang}, \citenamefont {Qian}, \citenamefont {Barin}, \citenamefont {Daaoub}, \citenamefont {Chen}, \citenamefont {M{\"u}llen}, \citenamefont {Sangtarash}, \citenamefont {Ruffieux}, \citenamefont {Fasel}, \citenamefont {Sadeghi} \emph {et~al.}}]{ZQB23}%
  \BibitemOpen
  \bibfield  {author} {\bibinfo {author} {\bibfnamefont {Jian}\ \bibnamefont {Zhang}}, \bibinfo {author} {\bibfnamefont {Liu}\ \bibnamefont {Qian}}, \bibinfo {author} {\bibfnamefont {Gabriela~Borin}\ \bibnamefont {Barin}}, \bibinfo {author} {\bibfnamefont {Abdalghani~HS}\ \bibnamefont {Daaoub}}, \bibinfo {author} {\bibfnamefont {Peipei}\ \bibnamefont {Chen}}, \bibinfo {author} {\bibfnamefont {Klaus}\ \bibnamefont {M{\"u}llen}}, \bibinfo {author} {\bibfnamefont {Sara}\ \bibnamefont {Sangtarash}}, \bibinfo {author} {\bibfnamefont {Pascal}\ \bibnamefont {Ruffieux}}, \bibinfo {author} {\bibfnamefont {Roman}\ \bibnamefont {Fasel}}, \bibinfo {author} {\bibfnamefont {Hatef}\ \bibnamefont {Sadeghi}},  \emph {et~al.},\ }\bibfield  {title} {\enquote {\bibinfo {title} {Contacting individual graphene nanoribbons using carbon nanotube electrodes},}\ }\href {\doibase 10.1038/s41928-023-00991-3} {\bibfield  {journal} {\bibinfo  {journal} {Nature electronics}\ }\textbf {\bibinfo {volume} {6}},\ \bibinfo {pages} {572--581}
  (\bibinfo {year} {2023})}\BibitemShut {NoStop}%
\bibitem [{\citenamefont {Kim}\ \emph {et~al.}(2016)\citenamefont {Kim}, \citenamefont {Jang}, \citenamefont {Brar}, \citenamefont {Tolstova}, \citenamefont {Mauser},\ and\ \citenamefont {Atwater}}]{KJB16}%
  \BibitemOpen
  \bibfield  {author} {\bibinfo {author} {\bibfnamefont {Seyoon}\ \bibnamefont {Kim}}, \bibinfo {author} {\bibfnamefont {Min~Seok}\ \bibnamefont {Jang}}, \bibinfo {author} {\bibfnamefont {Victor~W.}\ \bibnamefont {Brar}}, \bibinfo {author} {\bibfnamefont {Yulia}\ \bibnamefont {Tolstova}}, \bibinfo {author} {\bibfnamefont {Kelly~W.}\ \bibnamefont {Mauser}}, \ and\ \bibinfo {author} {\bibfnamefont {Harry~A.}\ \bibnamefont {Atwater}},\ }\bibfield  {title} {\enquote {\bibinfo {title} {Electronically tunable extraordinary optical transmission in graphene plasmonic ribbons coupled to subwavelength metallic slit arrays},}\ }\href@noop {} {\bibfield  {journal} {\bibinfo  {journal} {Nat.\ Commun.}\ }\textbf {\bibinfo {volume} {7}},\ \bibinfo {pages} {12323} (\bibinfo {year} {2016})}\BibitemShut {NoStop}%
\bibitem [{\citenamefont {Sisler}\ \emph {et~al.}(2024)\citenamefont {Sisler}, \citenamefont {Thureja}, \citenamefont {Grajower}, \citenamefont {Sokhoyan}, \citenamefont {Huang},\ and\ \citenamefont {Atwater}}]{STG24}%
  \BibitemOpen
  \bibfield  {author} {\bibinfo {author} {\bibfnamefont {Jared}\ \bibnamefont {Sisler}}, \bibinfo {author} {\bibfnamefont {Prachi}\ \bibnamefont {Thureja}}, \bibinfo {author} {\bibfnamefont {Meir~Y}\ \bibnamefont {Grajower}}, \bibinfo {author} {\bibfnamefont {Ruzan}\ \bibnamefont {Sokhoyan}}, \bibinfo {author} {\bibfnamefont {Ivy}\ \bibnamefont {Huang}}, \ and\ \bibinfo {author} {\bibfnamefont {Harry~A}\ \bibnamefont {Atwater}},\ }\bibfield  {title} {\enquote {\bibinfo {title} {Electrically tunable space--time metasurfaces at optical frequencies},}\ }\href {\doibase 10.1038/s41565-024-01728-9} {\bibfield  {journal} {\bibinfo  {journal} {Nature Nanotechnology}\ }\textbf {\bibinfo {volume} {19}},\ \bibinfo {pages} {1491--1498} (\bibinfo {year} {2024})}\BibitemShut {NoStop}%
\bibitem [{\citenamefont {{Garc\'{\i}a de Abajo}}(2010{\natexlab{b}})}]{paper149}%
  \BibitemOpen
  \bibfield  {author} {\bibinfo {author} {\bibfnamefont {F.~J.}\ \bibnamefont {{Garc\'{\i}a de Abajo}}},\ }\bibfield  {title} {\enquote {\bibinfo {title} {Optical excitations in electron microscopy},}\ }\href {\doibase 10.1103/RevModPhys.82.209} {\bibfield  {journal} {\bibinfo  {journal} {Rev.\ Mod.\ Phys.}\ }\textbf {\bibinfo {volume} {82}},\ \bibinfo {pages} {209--275} (\bibinfo {year} {2010}{\natexlab{b}})}\BibitemShut {NoStop}%
\bibitem [{\citenamefont {Yu}\ \emph {et~al.}(2017{\natexlab{a}})\citenamefont {Yu}, \citenamefont {Cox}, \citenamefont {Saavedra},\ and\ \citenamefont {{Garc\'{\i}a de Abajo}}}]{paper303}%
  \BibitemOpen
  \bibfield  {author} {\bibinfo {author} {\bibfnamefont {R.}~\bibnamefont {Yu}}, \bibinfo {author} {\bibfnamefont {J.~D.}\ \bibnamefont {Cox}}, \bibinfo {author} {\bibfnamefont {J.~R.~M.}\ \bibnamefont {Saavedra}}, \ and\ \bibinfo {author} {\bibfnamefont {F.~J.}\ \bibnamefont {{Garc\'{\i}a de Abajo}}},\ }\bibfield  {title} {\enquote {\bibinfo {title} {Analytical modeling of graphene plasmons},}\ }\href {\doibase 10.1021/acsphotonics.7b00740} {\bibfield  {journal} {\bibinfo  {journal} {ACS\ Photonics}\ }\textbf {\bibinfo {volume} {4}},\ \bibinfo {pages} {3106--3114} (\bibinfo {year} {2017}{\natexlab{a}})}\BibitemShut {NoStop}%
\bibitem [{\citenamefont {Yu}\ \emph {et~al.}(2017{\natexlab{b}})\citenamefont {Yu}, \citenamefont {Liz-Marz\'an},\ and\ \citenamefont {{Garc\'{\i}a de Abajo}}}]{paper300}%
  \BibitemOpen
  \bibfield  {author} {\bibinfo {author} {\bibfnamefont {R.}~\bibnamefont {Yu}}, \bibinfo {author} {\bibfnamefont {L.~M.}\ \bibnamefont {Liz-Marz\'an}}, \ and\ \bibinfo {author} {\bibfnamefont {F.~J.}\ \bibnamefont {{Garc\'{\i}a de Abajo}}},\ }\bibfield  {title} {\enquote {\bibinfo {title} {Universal analytical modeling of plasmonic nanoparticles},}\ }\href {\doibase 10.1039/c6cs00919k} {\bibfield  {journal} {\bibinfo  {journal} {Chem.\ Soc.\ Rev.}\ }\textbf {\bibinfo {volume} {46}},\ \bibinfo {pages} {6710--6724} (\bibinfo {year} {2017}{\natexlab{b}})}\BibitemShut {NoStop}%
\bibitem [{\citenamefont {Ashcroft}\ and\ \citenamefont {Mermin}(1976)}]{AM1976}%
  \BibitemOpen
  \bibfield  {author} {\bibinfo {author} {\bibfnamefont {N.~W.}\ \bibnamefont {Ashcroft}}\ and\ \bibinfo {author} {\bibfnamefont {N.~D.}\ \bibnamefont {Mermin}},\ }\href@noop {} {\emph {\bibinfo {title} {Solid State Physics}}}\ (\bibinfo  {publisher} {Harcourt College Publishers},\ \bibinfo {address} {Philadelphia},\ \bibinfo {year} {1976})\BibitemShut {NoStop}%
\bibitem [{\citenamefont {Bruggeman}(1935)}]{B1935}%
  \BibitemOpen
  \bibfield  {author} {\bibinfo {author} {\bibfnamefont {D.~A.~G.}\ \bibnamefont {Bruggeman}},\ }\bibfield  {title} {\enquote {\bibinfo {title} {Calculation of various physics constants in heterogenous substances. dielectricity constants and conductivity of mixed bodies from isotropic substances},}\ }\href@noop {} {\bibfield  {journal} {\bibinfo  {journal} {Ann.\ Phys.\ (Leipzig)}\ }\textbf {\bibinfo {volume} {24}},\ \bibinfo {pages} {636--664} (\bibinfo {year} {1935})}\BibitemShut {NoStop}%
\bibitem [{\citenamefont {Beaini}\ \emph {et~al.}(2020)\citenamefont {Beaini}, \citenamefont {Baloukas}, \citenamefont {Loquai}, \citenamefont {Klemberg-Sapieha},\ and\ \citenamefont {Martinu}}]{BBL20}%
  \BibitemOpen
  \bibfield  {author} {\bibinfo {author} {\bibfnamefont {R.}~\bibnamefont {Beaini}}, \bibinfo {author} {\bibfnamefont {B.}~\bibnamefont {Baloukas}}, \bibinfo {author} {\bibfnamefont {S.}~\bibnamefont {Loquai}}, \bibinfo {author} {\bibfnamefont {J.-E.}\ \bibnamefont {Klemberg-Sapieha}}, \ and\ \bibinfo {author} {\bibfnamefont {L.}~\bibnamefont {Martinu}},\ }\bibfield  {title} {\enquote {\bibinfo {title} {Thermochromic vo$_2$-based smart radiator devices with ultralow refractive index cavities for increased performance},}\ }\href {\doibase 10.1016/j.solmat.2020.110260} {\bibfield  {journal} {\bibinfo  {journal} {Solar Energy Materials and Solar Cells}\ }\textbf {\bibinfo {volume} {205}},\ \bibinfo {pages} {110260} (\bibinfo {year} {2020})}\BibitemShut {NoStop}%
\bibitem [{\citenamefont {Gon\c{c}alves}\ and\ \citenamefont {Peres}(2016)}]{GP16}%
  \BibitemOpen
  \bibfield  {author} {\bibinfo {author} {\bibfnamefont {P.~A.~D.}\ \bibnamefont {Gon\c{c}alves}}\ and\ \bibinfo {author} {\bibfnamefont {N.~M.~R.}\ \bibnamefont {Peres}},\ }\href@noop {} {\emph {\bibinfo {title} {An Introduction to Graphene Plasmonics}}}\ (\bibinfo  {publisher} {World Scientific},\ \bibinfo {address} {Singapore},\ \bibinfo {year} {2016})\BibitemShut {NoStop}%
\bibitem [{\citenamefont {Johnson}\ and\ \citenamefont {Christy}(1972)}]{JC1972}%
  \BibitemOpen
  \bibfield  {author} {\bibinfo {author} {\bibfnamefont {P.~B.}\ \bibnamefont {Johnson}}\ and\ \bibinfo {author} {\bibfnamefont {R.~W.}\ \bibnamefont {Christy}},\ }\bibfield  {title} {\enquote {\bibinfo {title} {Optical constants of the noble metals},}\ }\href {\doibase 10.1103/PhysRevB.6.4370} {\bibfield  {journal} {\bibinfo  {journal} {Phys.\ Rev.\ B}\ }\textbf {\bibinfo {volume} {6}},\ \bibinfo {pages} {4370--4379} (\bibinfo {year} {1972})}\BibitemShut {NoStop}%
\bibitem [{\citenamefont {{Garc\'{\i}a de Abajo}}(2013)}]{paper228}%
  \BibitemOpen
  \bibfield  {author} {\bibinfo {author} {\bibfnamefont {F.~J.}\ \bibnamefont {{Garc\'{\i}a de Abajo}}},\ }\bibfield  {title} {\enquote {\bibinfo {title} {Multiple excitation of confined graphene plasmons by single free electrons},}\ }\href {\doibase 10.1021/nn405367e} {\bibfield  {journal} {\bibinfo  {journal} {ACS\ Nano}\ }\textbf {\bibinfo {volume} {7}},\ \bibinfo {pages} {11409--11419} (\bibinfo {year} {2013})}\BibitemShut {NoStop}%
\bibitem [{\citenamefont {Rasmussen}\ \emph {et~al.}(2024)\citenamefont {Rasmussen}, \citenamefont {{Rodríguez Echarri}}, \citenamefont {Cox},\ and\ \citenamefont {{Garc\'{\i}a de Abajo}}}]{paper430}%
  \BibitemOpen
  \bibfield  {author} {\bibinfo {author} {\bibfnamefont {T.~P.}\ \bibnamefont {Rasmussen}}, \bibinfo {author} {\bibfnamefont {A.}~\bibnamefont {{Rodríguez Echarri}}}, \bibinfo {author} {\bibfnamefont {J.~D.}\ \bibnamefont {Cox}}, \ and\ \bibinfo {author} {\bibfnamefont {F.~J.}\ \bibnamefont {{Garc\'{\i}a de Abajo}}},\ }\bibfield  {title} {\enquote {\bibinfo {title} {Generation of entangled waveguided photon pairs by free electrons},}\ }\href {\doibase 10.1126/sciadv.eadn6312} {\bibfield  {journal} {\bibinfo  {journal} {Sci.\ Adv.}\ }\textbf {\bibinfo {volume} {10}},\ \bibinfo {pages} {eadn6312} (\bibinfo {year} {2024})}\BibitemShut {NoStop}%
\bibitem [{\citenamefont {Gomez-Heredia}\ \emph {et~al.}(2018)\citenamefont {Gomez-Heredia}, \citenamefont {Ramirez-Rincon}, \citenamefont {Ordonez-Miranda}, \citenamefont {Ares}, \citenamefont {Alvarado-Gil}, \citenamefont {Champeaux}, \citenamefont {Dumas-Bouchiat}, \citenamefont {Ezzahri},\ and\ \citenamefont {Joulain}}]{GRO18}%
  \BibitemOpen
  \bibfield  {author} {\bibinfo {author} {\bibfnamefont {C.~L.}\ \bibnamefont {Gomez-Heredia}}, \bibinfo {author} {\bibfnamefont {J.~A.}\ \bibnamefont {Ramirez-Rincon}}, \bibinfo {author} {\bibfnamefont {J.}~\bibnamefont {Ordonez-Miranda}}, \bibinfo {author} {\bibfnamefont {O.}~\bibnamefont {Ares}}, \bibinfo {author} {\bibfnamefont {J.~J.}\ \bibnamefont {Alvarado-Gil}}, \bibinfo {author} {\bibfnamefont {C.}~\bibnamefont {Champeaux}}, \bibinfo {author} {\bibfnamefont {F.}~\bibnamefont {Dumas-Bouchiat}}, \bibinfo {author} {\bibfnamefont {Y.}~\bibnamefont {Ezzahri}}, \ and\ \bibinfo {author} {\bibfnamefont {K.}~\bibnamefont {Joulain}},\ }\bibfield  {title} {\enquote {\bibinfo {title} {Thermal hysteresis measurement of the vo$_2$ emissivity and its application in thermal rectification},}\ }\href {\doibase 10.1038/s41598-018-26687-9} {\bibfield  {journal} {\bibinfo  {journal} {Scientific Reports}\ }\textbf {\bibinfo {volume} {8}},\ \bibinfo {pages} {8479} (\bibinfo {year} {2018})}\BibitemShut {NoStop}%
\bibitem [{\citenamefont {Hamaoui}\ \emph {et~al.}(2019)\citenamefont {Hamaoui}, \citenamefont {Horny}, \citenamefont {Gomez-Heredia}, \citenamefont {Ramirez-Rincon}, \citenamefont {Chirtoc}, \citenamefont {Ezzahri},\ and\ \citenamefont {Joulain}}]{HHG19}%
  \BibitemOpen
  \bibfield  {author} {\bibinfo {author} {\bibfnamefont {Georges}\ \bibnamefont {Hamaoui}}, \bibinfo {author} {\bibfnamefont {Nicolas}\ \bibnamefont {Horny}}, \bibinfo {author} {\bibfnamefont {Cindy~Lorena}\ \bibnamefont {Gomez-Heredia}}, \bibinfo {author} {\bibfnamefont {J.~A.}\ \bibnamefont {Ramirez-Rincon}}, \bibinfo {author} {\bibfnamefont {Mihai}\ \bibnamefont {Chirtoc}}, \bibinfo {author} {\bibfnamefont {Youn{\`e}s}\ \bibnamefont {Ezzahri}}, \ and\ \bibinfo {author} {\bibfnamefont {Karl}\ \bibnamefont {Joulain}},\ }\bibfield  {title} {\enquote {\bibinfo {title} {Thermophysical characterisation of vo$_2$ thin films: hysteresis and its application in thermal rectification},}\ }\href {\doibase 10.1038/s41598-019-45436-0} {\bibfield  {journal} {\bibinfo  {journal} {Scientific Reports}\ }\textbf {\bibinfo {volume} {9}},\ \bibinfo {pages} {8728} (\bibinfo {year} {2019})}\BibitemShut {NoStop}%
\bibitem [{\citenamefont {Voller}\ and\ \citenamefont {Cross}(1981)}]{VC81}%
  \BibitemOpen
  \bibfield  {author} {\bibinfo {author} {\bibfnamefont {V.~R.}\ \bibnamefont {Voller}}\ and\ \bibinfo {author} {\bibfnamefont {M.}~\bibnamefont {Cross}},\ }\bibfield  {title} {\enquote {\bibinfo {title} {Accurate solutions of moving boundary problems using the enthalpy method},}\ }\href {\doibase 10.1016/0017-9310(81)90062-4} {\bibfield  {journal} {\bibinfo  {journal} {International Journal of Heat and Mass Transfer}\ }\textbf {\bibinfo {volume} {24}},\ \bibinfo {pages} {545--556} (\bibinfo {year} {1981})}\BibitemShut {NoStop}%
\bibitem [{\citenamefont {Oh}\ \emph {et~al.}(2010)\citenamefont {Oh}, \citenamefont {Ko}, \citenamefont {Ramanathan},\ and\ \citenamefont {Cahill}}]{OKR10}%
  \BibitemOpen
  \bibfield  {author} {\bibinfo {author} {\bibfnamefont {Dong-Wook}\ \bibnamefont {Oh}}, \bibinfo {author} {\bibfnamefont {Changhyun}\ \bibnamefont {Ko}}, \bibinfo {author} {\bibfnamefont {Shriram}\ \bibnamefont {Ramanathan}}, \ and\ \bibinfo {author} {\bibfnamefont {David~G.}\ \bibnamefont {Cahill}},\ }\bibfield  {title} {\enquote {\bibinfo {title} {Thermal conductivity and dynamic heat capacity across the metal-insulator transition in thin film vo$_2$},}\ }\href {\doibase 10.1063/1.3394016} {\bibfield  {journal} {\bibinfo  {journal} {Applied Physics Letters}\ }\textbf {\bibinfo {volume} {96}},\ \bibinfo {pages} {151906} (\bibinfo {year} {2010})}\BibitemShut {NoStop}%
\bibitem [{\citenamefont {Avrami}(1939)}]{A1939}%
  \BibitemOpen
  \bibfield  {author} {\bibinfo {author} {\bibfnamefont {Melvin}\ \bibnamefont {Avrami}},\ }\bibfield  {title} {\enquote {\bibinfo {title} {Kinetics of phase change. {I}. general theory},}\ }\href {\doibase 10.1063/1.1750380} {\bibfield  {journal} {\bibinfo  {journal} {The Journal of Chemical Physics}\ }\textbf {\bibinfo {volume} {7}},\ \bibinfo {pages} {1103--1112} (\bibinfo {year} {1939})}\BibitemShut {NoStop}%
\bibitem [{\citenamefont {Farjas}\ and\ \citenamefont {Roura}(2006)}]{FR06}%
  \BibitemOpen
  \bibfield  {author} {\bibinfo {author} {\bibfnamefont {Jordi}\ \bibnamefont {Farjas}}\ and\ \bibinfo {author} {\bibfnamefont {Pere}\ \bibnamefont {Roura}},\ }\bibfield  {title} {\enquote {\bibinfo {title} {Modification of the {Kolmogorov--Johnson--Mehl--Avrami} rate equation for non-isothermal experiments and its analytical solution},}\ }\href {\doibase 10.1016/j.actamat.2006.07.037} {\bibfield  {journal} {\bibinfo  {journal} {Acta Materialia}\ }\textbf {\bibinfo {volume} {54}},\ \bibinfo {pages} {5573--5579} (\bibinfo {year} {2006})}\BibitemShut {NoStop}%
\bibitem [{\citenamefont {Chen}\ \emph {et~al.}(2022)\citenamefont {Chen}, \citenamefont {Han}, \citenamefont {Kong}, \citenamefont {Wang}, \citenamefont {Gu}, \citenamefont {Gao}, \citenamefont {Wang},\ and\ \citenamefont {Shen}}]{CHK22}%
  \BibitemOpen
  \bibfield  {author} {\bibinfo {author} {\bibfnamefont {Yimin}\ \bibnamefont {Chen}}, \bibinfo {author} {\bibfnamefont {Nan}\ \bibnamefont {Han}}, \bibinfo {author} {\bibfnamefont {Fanshuo}\ \bibnamefont {Kong}}, \bibinfo {author} {\bibfnamefont {Jun-Qiang}\ \bibnamefont {Wang}}, \bibinfo {author} {\bibfnamefont {Chenjie}\ \bibnamefont {Gu}}, \bibinfo {author} {\bibfnamefont {Yixiao}\ \bibnamefont {Gao}}, \bibinfo {author} {\bibfnamefont {Guoxiang}\ \bibnamefont {Wang}}, \ and\ \bibinfo {author} {\bibfnamefont {Xiang}\ \bibnamefont {Shen}},\ }\bibfield  {title} {\enquote {\bibinfo {title} {Kinetics features of 2d confined {Ge$_2$Sb$_2$Te$_5$} ultrathin film},}\ }\href {\doibase 10.1063/5.0100570} {\bibfield  {journal} {\bibinfo  {journal} {Applied Physics Letters}\ }\textbf {\bibinfo {volume} {121}},\ \bibinfo {pages} {061904} (\bibinfo {year} {2022})}\BibitemShut {NoStop}%
\bibitem [{\citenamefont {Kumar}\ \emph {et~al.}(2021)\citenamefont {Kumar}, \citenamefont {Shaikh},\ and\ \citenamefont {Chuang}}]{KSC21}%
  \BibitemOpen
  \bibfield  {author} {\bibinfo {author} {\bibfnamefont {Amit}\ \bibnamefont {Kumar}}, \bibinfo {author} {\bibfnamefont {Muhammad~Omar}\ \bibnamefont {Shaikh}}, \ and\ \bibinfo {author} {\bibfnamefont {Cheng-Hsin}\ \bibnamefont {Chuang}},\ }\bibfield  {title} {\enquote {\bibinfo {title} {Silver nanowire synthesis and strategies for fabricating transparent conducting electrodes},}\ }\href {\doibase 10.3390/nano11030693} {\bibfield  {journal} {\bibinfo  {journal} {Nanomaterials}\ }\textbf {\bibinfo {volume} {11}},\ \bibinfo {pages} {693} (\bibinfo {year} {2021})}\BibitemShut {NoStop}%
\bibitem [{\citenamefont {Pan}\ \emph {et~al.}(2024)\citenamefont {Pan}, \citenamefont {Tong}, \citenamefont {Qian}, \citenamefont {Krasavin}, \citenamefont {Li}, \citenamefont {Zhu}, \citenamefont {Zhang}, \citenamefont {Cui}, \citenamefont {Li}, \citenamefont {Wu} \emph {et~al.}}]{PTQ24}%
  \BibitemOpen
  \bibfield  {author} {\bibinfo {author} {\bibfnamefont {Chenxinyu}\ \bibnamefont {Pan}}, \bibinfo {author} {\bibfnamefont {Yuanbiao}\ \bibnamefont {Tong}}, \bibinfo {author} {\bibfnamefont {Haoliang}\ \bibnamefont {Qian}}, \bibinfo {author} {\bibfnamefont {Alexey~V}\ \bibnamefont {Krasavin}}, \bibinfo {author} {\bibfnamefont {Jialin}\ \bibnamefont {Li}}, \bibinfo {author} {\bibfnamefont {Jiajie}\ \bibnamefont {Zhu}}, \bibinfo {author} {\bibfnamefont {Yiyun}\ \bibnamefont {Zhang}}, \bibinfo {author} {\bibfnamefont {Bowen}\ \bibnamefont {Cui}}, \bibinfo {author} {\bibfnamefont {Zhiyong}\ \bibnamefont {Li}}, \bibinfo {author} {\bibfnamefont {Chenming}\ \bibnamefont {Wu}},  \emph {et~al.},\ }\bibfield  {title} {\enquote {\bibinfo {title} {Large area single crystal gold of single nanometer thickness for nanophotonics},}\ }\href {\doibase 10.1038/s41467-024-47133-7} {\bibfield  {journal} {\bibinfo  {journal} {Nature Communications}\ }\textbf {\bibinfo {volume} {15}},\ \bibinfo {pages} {2840} (\bibinfo {year}
  {2024})}\BibitemShut {NoStop}%
\end{thebibliography}
\end{document}